\documentclass[ALICE,manyauthors]{cernphprep}
\usepackage[comma,square,numbers,sort&compress]{natbib}
\usepackage{hyperref}
\usepackage{lineno}
\usepackage{xspace}
\usepackage{float}
\usepackage{epstopdf}
\usepackage{upgreek} % For upright Greek letters
\usepackage{rotating}
\usepackage[T1]{fontenc}
\usepackage{orcidlink}

\begin{document}
%%%%%%%%%%%%%%%%%%%%%%%%%%%%%%%%%%%%%%%%%%%%%%%%%%
% These are some new commands that may be useful 
% for paper writing in general. If other newcommands
% are needed for your specific paper, please feel 
% free to add here. 
%
% The currently available commands are organized in: 
% 1) Systems
% 2) Quantities
% 3) Energies and units
% 4) Detectors
% 5) particle species 
%%%%%%%%%%%%%%%%%%%%%%%%%%%%%%%%%%%%%%%%%%%%%%%%%%

% 1) SYSTEMS 
\newcommand{\pp}           {pp\xspace}
\newcommand{\ppbar}        {\mbox{$\mathrm {p\overline{p}}$}\xspace}
\newcommand{\XeXe}         {\mbox{Xe--Xe}\xspace}
\newcommand{\PbPb}         {\mbox{Pb--Pb}\xspace}
\newcommand{\pA}           {\mbox{pA}\xspace}
\newcommand{\pPb}          {\mbox{p--Pb}\xspace}
\newcommand{\AuAu}         {\mbox{Au--Au}\xspace}
\newcommand{\dAu}          {\mbox{d--Au}\xspace}

% 2) QUANTITIES 
\newcommand{\s}            {\ensuremath{\sqrt{s}}\xspace}
\newcommand{\snn}          {\ensuremath{\sqrt{s_{\mathrm{NN}}}}\xspace}
\newcommand{\pt}           {\ensuremath{p_{\rm T}}\xspace}
\newcommand{\meanpt}       {$\langle p_{\mathrm{T}}\rangle$\xspace}
\newcommand{\ycms}         {\ensuremath{y_{\rm CMS}}\xspace}
\newcommand{\ylab}         {\ensuremath{y_{\rm lab}}\xspace}
\newcommand{\etarange}[1]  {\mbox{$\left | \eta \right |~<~#1$}}
\newcommand{\yrange}[1]    {\mbox{$\left | y \right |~<~#1$}}
\newcommand{\dndy}         {\ensuremath{\mathrm{d}N_\mathrm{ch}/\mathrm{d}y}\xspace}
\newcommand{\dndeta}       {\ensuremath{\mathrm{d}N_\mathrm{ch}/\mathrm{d}\eta}\xspace}
\newcommand{\avdndeta}     {\ensuremath{\langle\dndeta\rangle}\xspace}
\newcommand{\dNdy}         {\ensuremath{\mathrm{d}N_\mathrm{ch}/\mathrm{d}y}\xspace}
\newcommand{\Npart}        {\ensuremath{N_\mathrm{part}}\xspace}
\newcommand{\Ncoll}        {\ensuremath{N_\mathrm{coll}}\xspace}
\newcommand{\dEdx}         {\ensuremath{\textrm{d}E/\textrm{d}x}\xspace}
\newcommand{\RpPb}         {\ensuremath{R_{\rm pPb}}\xspace}

% 3) ENERGIES, UNITS
\newcommand{\nineH}        {$\sqrt{s}~=~0.9$~Te\kern-.1emV\xspace}
\newcommand{\seven}        {$\sqrt{s}~=~7$~Te\kern-.1emV\xspace}
\newcommand{\twoH}         {$\sqrt{s}~=~0.2$~Te\kern-.1emV\xspace}
\newcommand{\twosevensix}  {$\sqrt{s}~=~2.76$~Te\kern-.1emV\xspace}
\newcommand{\five}         {$\sqrt{s}~=~5.02$~Te\kern-.1emV\xspace}
\newcommand{\twosevensixnn}{$\sqrt{s_{\mathrm{NN}}}~=~2.76$~Te\kern-.1emV\xspace}
\newcommand{\fivenn}       {$\sqrt{s_{\mathrm{NN}}}~=~5.02$~Te\kern-.1emV\xspace}
\newcommand{\LT}           {L{\'e}vy-Tsallis\xspace}
\newcommand{\GeVc}         {Ge\kern-.1emV/$c$\xspace}
\newcommand{\MeVc}         {Me\kern-.1emV/$c$\xspace}
\newcommand{\TeV}          {Te\kern-.1emV\xspace}
\newcommand{\GeV}          {Ge\kern-.1emV\xspace}
\newcommand{\MeV}          {Me\kern-.1emV\xspace}
\newcommand{\GeVmass}      {Ge\kern-.2emV/$c^2$\xspace}
\newcommand{\MeVmass}      {Me\kern-.2emV/$c^2$\xspace}
\newcommand{\lumi}         {\ensuremath{\mathcal{L}}\xspace}

% 4) DETECTORS 
\newcommand{\ITS}          {\rm{ITS}\xspace}
\newcommand{\TOF}          {\rm{TOF}\xspace}
\newcommand{\ZDC}          {\rm{ZDC}\xspace}
\newcommand{\ZDCs}         {\rm{ZDCs}\xspace}
\newcommand{\ZNA}          {\rm{ZNA}\xspace}
\newcommand{\ZNC}          {\rm{ZNC}\xspace}
\newcommand{\SPD}          {\rm{SPD}\xspace}
\newcommand{\SDD}          {\rm{SDD}\xspace}
\newcommand{\SSD}          {\rm{SSD}\xspace}
\newcommand{\TPC}          {\rm{TPC}\xspace}
\newcommand{\TRD}          {\rm{TRD}\xspace}
\newcommand{\VZERO}        {\rm{V0}\xspace}
\newcommand{\VZEROA}       {\rm{V0A}\xspace}
\newcommand{\VZEROC}       {\rm{V0C}\xspace}
\newcommand{\Vdecay} 	   {\ensuremath{V^{0}}\xspace}

% 4) PARTICLE SPECIES 
\newcommand{\ee}           {\ensuremath{e^{+}e^{-}}} 
\newcommand{\pip}          {\ensuremath{\pi^{+}}\xspace}
\newcommand{\pim}          {\ensuremath{\pi^{-}}\xspace}
\newcommand{\kap}          {\ensuremath{\rm{K}^{+}}\xspace}
\newcommand{\kam}          {\ensuremath{\rm{K}^{-}}\xspace}
\newcommand{\pbar}         {\ensuremath{\rm\overline{p}}\xspace}
\newcommand{\kzero}        {\ensuremath{{\rm K}^{0}_{\rm{S}}}\xspace}
\newcommand{\lmb}          {\ensuremath{\Lambda}\xspace}
\newcommand{\almb}         {\ensuremath{\overline{\Lambda}}\xspace}
\newcommand{\Om}           {\ensuremath{\Omega^-}\xspace}
\newcommand{\Mo}           {\ensuremath{\overline{\Omega}^+}\xspace}
\newcommand{\X}            {\ensuremath{\Xi^-}\xspace}
\newcommand{\Ix}           {\ensuremath{\overline{\Xi}^+}\xspace}
\newcommand{\Xis}          {\ensuremath{\Xi^{\pm}}\xspace}
\newcommand{\Oms}          {\ensuremath{\Omega^{\pm}}\xspace}
\newcommand{\degree}       {\ensuremath{^{\rm o}}\xspace}

%%%%%%%%%%%%%%%  Title page %%%%%%%%%%%%%%%%%%%%%%%%
\begin{titlepage}
% the dates below correspond to CERN approval
% please don't touch: EB chairs will take care
\PHyear{2025}       % required, will be obtained from CERN
\PHnumber{066}      % required, will be obtained from CERN
\PHdate{20 March}  % required, will be obtained from CERN
%%%%%%%%%%%%%%%%%%%%%%%%%%%%%%%%%%%%%%%%%%%%%%%%%%%%

%%% Put your own title + short title here:
\title{Measurement of correlations among net-charge, net-proton, and net-kaon multiplicity distributions in Pb--Pb collisions at $\sqrt{s_\text{NN}}=5.02$~TeV}
\ShortTitle{Correlations among net-charged particle, net-proton and net-kaon}   % appears on left page headers

%%% Do not change the next lines
\Collaboration{ALICE Collaboration\thanks{See Appendix~\ref{app:collab} for the list of collaboration members}}
\ShortAuthor{ALICE Collaboration} % appears on right page headers, do not change

\begin{abstract}
Correlations among conserved quantum numbers, such as the net-electric charge, the net-baryon, and the net-strangeness in heavy-ion collisions, are crucial for exploring the QCD phase diagram. In this paper, these correlations are investigated using net-proton number (as a proxy for the net-baryon), net-kaon number (for the net-strangeness), and net-charged particle number in Pb--Pb collisions at $\sqrt{s_\text{NN}}=5.02$~TeV with the ALICE detector. The observed correlations deviate from the Poissonian baseline, with a more pronounced deviation at LHC energies than at RHIC. Theoretical calculations of the Thermal-FIST hadron resonance gas model, HIJING, and EPOS LHC event generators are compared with experimental results, where a significant impact of resonance decays is observed. Thermal-FIST calculations under the grand canonical and canonical ensembles highlight significant differences, underscoring the role of local charge conservation in explaining the data. Recent lattice QCD studies have demonstrated that the magnetic field generated by spectator protons in heavy-ion collisions affects susceptibility ratios, in particular those related to the net-electric charge and the net-baryon numbers. The experimental findings are in qualitative agreement with the expectations of lattice QCD. 

\end{abstract}
\end{titlepage}

\setcounter{page}{2} %please do not remove this line

%%%%%%%%%%%%%%%%%%%%%%%%%%%%%%%%
% begin main text
%%%%%%%%%%%%%%%%%%%%%%%%%%%%%%%%

\section{Introduction}
Fluctuations and correlations among conserved charges, namely the net-electric charge (Q), the net-baryon (B), and the net-strangeness (S), reveal key properties of strongly interacting matter~\cite{Shuryak:1997yj, Stephanov:1999zu, Heiselberg:2000fk, ALICE:2022wpn, Koch:2005vg, Gavai:2005yk, Bluhm:2008sc, Hietanen:2008xb}. These conserved charges are defined as the event-by-event difference between the number of particles and antiparticles carrying electric charge, baryon number, or strangeness in the system. Studying them offers insights into the phase structure of quantum chromodynamics (QCD), as they are accessible through both heavy-ion collision experiments and first-principle lattice QCD (LQCD) calculations via thermodynamic susceptibilities~\cite{Gupta:2011wh, HotQCD:2012fhj, Ding:2015ona, HotQCD:2019xnw, Borsanyi:2018grb}. At LHC energies, where the baryon chemical potential ($\mu_\mathrm{B}$) is close to zero~\cite{ALICE:2023ulv}, LQCD predicts that the transition from the low-temperature hadronic matter to the high-temperature quark$-$gluon plasma (QGP) occurs as a smooth crossover~\cite{Aoki:2006we}. At larger values of $\mu_\mathrm{B}$, a first-order phase transition line is anticipated~\cite{Ejiri:2008xt}. The first-order phase transition line is expected to end at a critical point (CP)~\cite{Berges:1998sd, Asakawa:1989bq, Stephanov:2006zvm}, near which the thermodynamic susceptibilities show a divergent behavior~\cite{Stephanov:2008qz, Asakawa:2009aj, Stephanov:2011pb}. Determining the precise location of the CP remains an active area of research, pursued through both experimental measurements and theoretical studies. Fluctuations and correlations of these conserved quantities serve as sensitive probes for analyzing freeze-out conditions in heavy-ion collisions and may help elucidate their connection to the QCD phase transition~\cite{Karsch:2010ck, Friman:2011pf}. Active experimental programs, such as the RHIC Beam Energy Scan (BES) and the LHC, are extensively investigating these using cumulant analyses~\cite{STAR:2020tga, STAR:2017tfy, Luo:2017faz, Pandav:2022xxx, ALICE:2022xpf}.  

The cumulants ($\kappa_{\alpha,\beta}^{lm}$) of net-multiplicity distributions for associated conserved charges ($\alpha$, $\beta$) can be arranged in a matrix form, where diagonal cumulants quantify fluctuations of individual conserved charges, and off-diagonal cumulants capture correlations between different conserved charges. They are related to the corresponding thermodynamic susceptibilities ($\chi_{\alpha,\beta}^{lm}$) by the equation:
\begin{equation}
  \chi_\mathrm{\alpha,\beta}^{lm}=\frac{1}{VT^{3}}\kappa_\mathrm{\alpha,\beta}^{lm}\,,
\end{equation}
where $V$ and $T$ denote the volume and temperature of the thermodynamic system. This study focuses on second-order fluctuations and correlations, with $\kappa_\mathrm{\alpha}^{2}$ and $\kappa_\mathrm{\alpha,\beta}^{11}$ measured using the relations:
\begin{equation}
  \label{eq:diagcumulant}
  \kappa_\mathrm{\alpha}^{2}=\langle(\delta N_\alpha)^{2}\rangle\,,
\end{equation}
and
\begin{equation}
   \label{eq:offdiagcumulant}
  \kappa_\mathrm{\alpha,\beta}^{11}=\langle(\delta N_\alpha)(\delta N_\beta)\rangle\,,
\end{equation}
where $\delta N_\alpha=(N_{\alpha^+}-N_{\alpha^-})-\langle(N_{\alpha^+}-N_{\alpha^-})\rangle$ and angular brackets $\langle ...\rangle$ denote average over all events. Here, $N_{\alpha^+}$ and $N_{\alpha^-}$ represent the event-by-event numbers of particles and antiparticles associated with the conserved charge $\alpha$, respectively. The use of cumulant ratios eliminates the temperature and volume dependence and enables a more direct comparison between experimental results and theoretical models. Since not all baryons and strange particles can be measured experimentally, the net-proton (difference between number of protons and antiprotons, calculated as $\Delta\mathrm{p} = N_{\mathrm{p}} - N_{\bar{\mathrm{p}}}$) and net-kaon (calculated as $\Delta\mathrm{K} = N_{\mathrm{K}^{+}} - N_{\mathrm{K}^{-}}$) numbers are used as proxies for the net-baryon and the net-strangeness numbers, respectively. The net-charged particle ($\Delta\mathrm{Q}$) is defined as the sum of net-pion (calculated as $\Delta\mathrm{\pi} = N_{\pi^{+}} - N_{\pi^{-}}$), net-kaon, and net-proton, i.e., $\Delta\mathrm{Q}=\Delta\pi+\Delta\mathrm{K}+\Delta\mathrm{p}$. The impact of other hadron species on the diagonal and off-diagonal susceptibilities of conserved charges (Q, B, and S) has been investigated with the Hadron Resonance Gas (HRG) model~\cite{Bellwied:2019pxh}. These studies show that the baryon-electric charge (BQ) and electric charge-strangeness (QS) correlations are dominated by contributions from protons and kaons, which, along with pions, form the bulk of particle production in heavy-ion collisions. In contrast, the baryon-strangeness (BS) correlation receives significant contributions from strange baryons, including hyperons and multi-strange baryons, which are often difficult to measure on an event-by-event basis due to limited reconstruction efficiencies. The importance of investigating the off-diagonal cumulants was highlighted in Ref.~\cite{Koch:2005vg} through studies of BS correlations; $C_\mathrm{B,S} = -3\chi_\mathrm{B,S}^{11}/\chi_\mathrm{S}^{2}$, referred to as ``Koch ratio''. The STAR experiment at RHIC has reported correlations among net-charged particle, net-proton, and net-kaon, i.e., $C_\mathrm{p,K}=\kappa_\mathrm{p,K}^{11}/\kappa_\mathrm{K}^{2}$, $C_\mathrm{Q,K}=\kappa_\mathrm{Q,K}^{11}/\kappa_\mathrm{K}^{2}$, and $C_\mathrm{Q,p}=\kappa_\mathrm{Q,p}^{11}/\kappa_\mathrm{p}^{2}$ in Au+Au collisions at $\sqrt{s_\mathrm{NN}}=7.7$ to 200 GeV~\cite{STAR:2019ans}. Since the off-diagonal cumulants $\kappa^{11}_\mathrm{Q,K}$ and $\kappa^{11}_\mathrm{Q,p}$ inherently include the diagonal cumulants of net-kaons and net-protons respectively, the ratios $C_\mathrm{Q,K}$ and $C_\mathrm{Q,p}$ can be expanded as:
\begin{equation}
C_\mathrm{Q,K}=\frac{\kappa^{11}_\mathrm{p,K}}{\kappa^{2}_\mathrm{K}}+\frac{\kappa^{11}_\mathrm{\pi,K}}{\kappa^{2}_\mathrm{K}}+1,
\end{equation}
and
\begin{equation}
C_\mathrm{Q,p}=\frac{\kappa^{11}_\mathrm{p,K}}{\kappa^{2}_\mathrm{p}}+\frac{\kappa^{11}_\mathrm{\pi,p}}{\kappa^{2}_\mathrm{p}}+1.
\end{equation}
At LHC energies, the off-diagonal cumulants are directly related to the balance function (BF) integrals~\cite{ALICE:2015nuz, ALICE:2021hjb}, which characterize correlations between oppositely charged particle pairs. Moreover, these cumulants are sensitive to several dynamical effects that can obscure the critical fluctuations of interest. Such effects include correlations from global and local conservation of quantum numbers, meaning charge conservation applied to the overall system as well as within localized regions~\cite{Vovchenko:2020kwg, Braun-Munzinger:2019yxj}, volume fluctuations arising from event-by-event changes in the system size, and resonance decays~\cite{Bluhm:2020mpc, Mishra:2016qyj} where unstable hadrons decay into stable particles. Additional contributions include initial-state fluctuations, which refer to random variations in the positions of nucleons within the colliding nuclei~\cite{Shen:2017bsr}, and thermal blurring, which reflects the smearing of measured fluctuations due to thermal motion of particles.~\cite{Ohnishi:2016bdf}. Thus, isolating and understanding the specific contributions of these fluctuations and correlations is crucial to make accurate inferences about the QCD phase structure and its critical behaviors.

In high-energy collisions, electric charge, baryon number, and strangeness are conserved in each event when all produced particles are considered. Consequently, fluctuations in the net-conserved charges vanish over the full phase space due to global conservation laws and must instead be studied within a restricted phase space, which can be achieved by selecting specific ranges in rapidity ($y$) and/or transverse momentum ($p_\mathrm{T}$) of the detected particles~\cite{Bzdak:2012an}. Although restricting the phase space suppresses global conservation effects, the selected window should not be smaller than the system's intrinsic dynamical correlation length, as this may mask the fluctuations of interest~\cite{Braun-Munzinger:2016yjz}. The acceptance fraction ($\alpha_\mathrm{acc}$), defined as the ratio of the phase space covered by kinematic cuts to the system's total phase space~\cite{Braun-Munzinger:2016yjz, Vovchenko:2020tsr, Vovchenko:2020gne}, plays an important role in understanding how the measured fluctuations and correlations relate to the theoretical calculations of grand-canonical susceptibilities within HRG models and LQCD calculations. Moreover, event-by-event fluctuations in net-particle numbers can be influenced by fluctuations in the number of participating nucleons from the projectile and target for a given centrality selection~\cite{WA98:2000mvt}. The effect of participant fluctuations (also known as volume fluctuations) on cumulants of the net-conserved charges has been studied within the framework of the Wounded Nucleon Model~\cite{Bialas:1976ed} in Ref.~\cite{Braun-Munzinger:2016yjz}. It was found that the cumulants are significantly affected by volume fluctuations at lower energies. In contrast, at LHC energies, where the mean net-particle numbers vanish, the cumulants are independent of volume fluctuations up to third order.

In the early stages of high-energy heavy-ion collisions with non-zero impact parameters, a strong magnetic field is generated by the spectator nucleons (those nucleons that do not participate directly in the collision but continue along the beam direction). At LHC energies, the field strength is estimated to reach $eB \sim 15m_\pi^{2} \sim 10^{15}$ T~\cite{Skokov:2009qp}. This field is transient and decays rapidly with time~\cite{Kharzeev:2007jp, Voronyuk:2011jd}. However, a QGP medium with finite electrical conductivity can partially sustain it~\cite{Mohapatra:2011ku, Tuchin:2010gx}, making the decay rate dependent on the medium's conductivity, which remains experimentally unconstrained. Depending on the electrical conductivity and the formation time of the QGP medium created in heavy-ion collisions, the initial magnetic field may influence the motion of final-state particles. There is considerable interest in understanding how such intense magnetic fields influence the properties and dynamics of final-state hadrons. Recently, LQCD calculations have indicated that such a magnetic field can influence specific combinations of susceptibilities related to electric charge, baryon number, and strangeness, which can be studied experimentally in peripheral heavy-ion collisions~\cite{Ding:2021cwv}. In the absence of a magnetic field, up ($u$) and down ($d$) quarks exhibit isospin symmetry, i.e., they interact with the strong nuclear force in the same way. However, the introduction of a magnetic field breaks this symmetry because of the different electric charges of the quarks, leading to changes in the quark-level susceptibilities ($\chi^{2}_u$ and $\chi^{2}_d$), which alter the entire second-order susceptibility matrix for electric charge, baryon number, and strangeness~\cite{Ding:2021cwv}. To study the effect of the magnetic field in the later (hadronic) stages of heavy-ion collisions, examining the centrality dependence of quantities such as $(2\chi^{11}_\mathrm{Q,S}-\chi^{11}_\mathrm{B,S})/\chi^{2}_\mathrm{S}$, $(2\chi^{11}_\mathrm{Q,B}-\chi^{11}_\mathrm{B,S})/\chi^{2}_\mathrm{B}$, and $\chi^{11}_\mathrm{Q,B}/\chi^{2}_\mathrm{Q}$ has recently been proposed in Refs.~\cite{Ding:2021cwv, Ding:2023bft}. In particular, the scaled ratio $[\chi^{11}_\mathrm{Q,B}/\chi^{2}_\mathrm{Q}(eB)]/[\chi^{11}_\mathrm{Q,B}/\chi^{2}_\mathrm{Q}(eB=0)]$, which shows significant deviations from unity, may serve as a sensitive indicator of the magnetic field’s effect~\cite{Ding:2023bft}. Experimentally, these quantities are accessed via proxies $(2\kappa^{11}_\mathrm{Q,K}-\kappa^{11}_\mathrm{p,K})/\kappa^{2}_\mathrm{K}$, $(2\kappa^{11}_\mathrm{Q,p}-\kappa^{11}_\mathrm{p,K})/\kappa^{2}_\mathrm{p}$, and $(\kappa^{11}_\mathrm{Q,p}/\kappa^{2}_\mathrm{Q})/(\kappa^{11}_\mathrm{Q,p}/\kappa^{2}_\mathrm{Q})^{0-5\%}$. The latter quantity $(\kappa^{11}_\mathrm{Q,p}/\kappa^{2}_\mathrm{Q})/(\kappa^{11}_\mathrm{Q,p}/\kappa^{2}_\mathrm{Q})^{0-5\%}$ represents the ratio of $\kappa^{11}_\mathrm{Q,p}/\kappa^{2}_\mathrm{Q}$ to its value in the 0$-$5\% centrality interval.

 The article is organized as follows. Section~\ref{sec:expsetup} provides a concise overview of the ALICE detector setup, focusing on the sub-detectors that are essential for this analysis. Details about the dataset and analysis \textemdash{} including event and track selection criteria, particle identification, the efficiency correction procedure, and the estimation of
statistical and systematic uncertainties, are also elaborated. The discussion on measurements and their comparison with HIJING and EPOS LHC event generators, as well as an HRG model Thermal-FIST can be found in Sec.~\ref{sec:results}. Finally, Sec.~\ref{sec:summary} summarizes the major findings of this analysis.

\section{Experimental setup and data analysis}
\label{sec:expsetup}
A comprehensive overview of the ALICE apparatus is available in Refs.~\cite{ALICE:2008ngc, ALICE:2014sbx}. The relevant detectors used in this analysis are the Inner Tracking System (ITS)~\cite{ALICE:2010tia}, a silicon detector comprising 6 cylindrical layers positioned near the collision point, the Time Projection Chamber (TPC)~\cite{Alme:2010ke}, a gaseous detector serving as ALICE's primary tracking and particle identification (PID) detector, and the Time-Of-Flight (TOF)~\cite{ALICE:2000xcm} detector, a gaseous type parallel plate chamber used for PID in the intermediate momentum range. Lastly, the V0 detector~\cite{ALICE:2013axi, ALICE:2004ftm}, consisting of two arrays (named V0A and V0C) of 32 scintillator tiles, is employed for triggering, event selection, and centrality determination. These arrays are located on either side of the interaction point, covering the pseudorapidity intervals $-3.7 < \eta < -1.6$ and $2.8 < \eta < 5.1$, respectively. Collision events are chosen using a minimum-bias (MB) trigger, requiring hits in both the V0A and V0C scintillators. Additionally, events with the primary vertex positioned within 10 cm relative to the nominal interaction point along the beam axis are exclusively selected to leverage the full detector acceptance. To ensure optimal detector performance, events with more than one reconstructed primary interaction vertex, referred to as pileup events~\cite{Arslandok:2022dyb}, are excluded. A total of approximately 80 million MB events are selected for the analysis. These events are categorized into centrality intervals based on the amplitude distribution measured in the V0 detector, corresponding to hadronic interactions as described in Ref.~\cite{ALICE:2013hur}. The measurements are performed in 18 centrality intervals from 0\% to 90\%, each with a width of 5\%. 

Nestled within a solenoid that generates a magnetic field of up to 0.5 T along the beam axis, ITS, TPC, and TOF provide complete azimuthal coverage for the charged particles in the pseudorapidity interval $|\eta| < 0.8$. Particles with at least one space point in the two innermost layers of the ITS and a minimum of 70 out of 159 space points in the TPC are selected for the analysis. To further refine the track selection and minimize contamination from secondary particles (i.e., particles originating from weak decays and material interactions), a $p_\mathrm{T}$-dependent selection criterion is applied to the distance of the closest approach to the primary vertex in the perpendicular plane ($\mathrm{DCA}_{xy}$) to the beam axis. Tracks are selected if their $\mathrm{DCA}_{xy}$ is less than $\mathrm{DCA}_{xy}(p_\mathrm{T}) = 0.0105+0.035/p_\mathrm{T}^{1.1}$ in cm. The largest allowed longitudinal distance of the tracks to the primary vertex ($\mathrm{DCA}_{z}$) is 2 cm. Additionally, the chi-square ($\chi^2$) per space point in the TPC and the ITS resulting from the track fit must be below 2.5 and 36, respectively. 
 
Pions ($\pi^+$), kaons ($\mathrm{K}^+$), and protons ($\mathrm{p}$) (as well as their antiparticles, $\pi^-$, $\mathrm{K}^-$, and $\bar{\mathrm{p}}$) are identified based on the specific energy loss $\mathrm{d}E/\mathrm{d}x$ in the TPC's gas volume and the flight time of a particle from the primary vertex of the collision to the TOF detector. The $p_\mathrm{T}$ ranges for $\pi^\pm$ and $\mathrm{K}^\pm$ are restricted to $0.2 < p_\mathrm{T} < 2.0$~GeV/$c$, while for p($\bar{\mathrm{p}}$), it is $0.4 < p_\mathrm{T} < 2.0$~GeV/$c$.  The variable $n(\sigma^\mathrm{TPC}_{i})$ quantifies the PID response in the TPC, representing the deviation between the measured and expected $\mathrm{d}E/\mathrm{d}x$ for a given particle species $i$, normalized by the detector resolution $\sigma(\mathrm{TPC})$. The expected $\mathrm{d}E/\mathrm{d}x$ is calculated using a parameterized Bethe-Bloch function~\cite{ALICE:2014sbx}. The $\pi^{\pm}$ and $\mathrm{K}^{\pm}$ are identified with $|n(\sigma^\mathrm{TPC}_i)|<2$ in the range $0.2 < p_\mathrm{T} < 0.5$~GeV/$\it{c}$ whereas the $\mathrm{p}$($\bar{\mathrm{p}}$) is selected using $|n(\sigma_{i}^\mathrm{TPC})|<2$ for $0.4 < p_\mathrm{T} < 0.6$~GeV/$\it{c}$. The lower limit for $p_\mathrm{T}$ is imposed because the reconstruction efficiency becomes very low and decreases rapidly below 0.2 GeV/$\it{c}$~\cite{ALICE:2013mez}. Additionally, a threshold of 0.4 GeV/$\it{c}$ is applied for protons to reduce contributions from protons generated by interactions of charged particles with the detector material.  Given that TPC's particle identification is limited to low momenta (where distinct $\mathrm{d}E/\mathrm{d}x$ bands are observed for different particle species), the TOF detector's information is also utilized for identifying tracks with $p_\mathrm{T}$ exceeding 0.5 GeV/$\it{c}$ (0.6 GeV/$\it{c}$ for protons). Similar to the TPC, $n(\sigma^\mathrm{TOF}_{i})$ represents the normalized difference between the measured and expected flight time of a particle species. Consequently, in the $0.5 < p_\mathrm{T} < 2.0$ GeV/$\it{c}$ range ($0.6 < p_\mathrm{T} < 2.0$ GeV/$\it{c}$ for protons), particle species are chosen if $|n(\sigma^\mathrm{TPC+TOF}_i)|<2$, where $n(\sigma^\mathrm{TPC+TOF}_i)=\sqrt{n(\sigma^\mathrm{TPC}_i)^2+n(\sigma^\mathrm{TOF}_i)^2}$. These selection criteria ensure that the integrated purity of the selected samples of $\pi^+$, $\mathrm{K}^+$, and $\mathrm{p}$ (and their antiparticles) remains above 98\% across the $p_\mathrm{T}$ ranges used in the analysis. The efficiencies for selecting the particles depend on both $p_\mathrm{T}$ and event centrality, ranging from 78\%, 50\%, and 80\% at $p_\mathrm{T} \sim$ 0.5 GeV/$c$ to 38\%, 30\%, and 40\% at $p_\mathrm{T} \sim$ 1.0 GeV/$c$ for $\pi^{\pm}$, $\mathrm{K}^{\pm}$, and p$(\bar{\mathrm{p}})$, respectively.

The results presented are based on data from Pb--Pb collisions at $\sqrt{s_\mathrm{NN}}=5.02$~TeV, collected by the ALICE collaboration during the Run 2 data-taking period of the LHC in 2015. The net-particle numbers for pion, kaon, and proton ($\Delta{\pi}$, $\Delta\mathrm{K}$, and $\Delta\mathrm{p}$) are measured event-by-event. Using these values, the associated cumulants (including both diagonal and off-diagonal) and the correlations $C_\mathrm{p,K}$, $C_\mathrm{Q,K}$, and $C_\mathrm{Q,p}$ are determined for different centrality intervals. The measurements are corrected for particle detection efficiency using the analytical correction method described in Refs.~\cite{Nonaka:2017kko, Bzdak:2013pha}, which assumes efficiency losses governed by the binomial statistics. The efficiency correction procedure was validated in two steps by a full Monte Carlo (MC) simulation using the HIJING~\cite{Wang:1991hta} event generator, where the generated particles are transported through the ALICE detector geometry modeled with GEANT4~\cite{Brun:1994aa} and reconstructed similarly to the collision data. First, the efficiencies for both particles and antiparticles of each species were determined independently as a function of $p_\mathrm{T}$ for each centrality interval. Then, the diagonal and off-diagonal cumulants obtained from the reconstructed MC tracks were corrected on a track-by-track basis using these efficiencies and the analytical expressions detailed in Refs.~\cite{Nonaka:2017kko, Luo:2018ofd}. An agreement of over 99\% was achieved between the efficiency-corrected results from the MC-reconstructed and those from the MC-generated data. 

\begin{table}[b]
  \begin{center}
      \caption{ \label{tab:sys} Contributions from various systematic uncertainty sources on $C_\mathrm{p,K}$, $C_\mathrm{Q,K}$, and $C_\mathrm{Q,p}$ in Pb--Pb collisions at $\sqrt{s_\mathrm{NN}}$ $=$ 5.02~TeV. The ranges correspond to the minimum and maximum values across different centrality interval.}

      \begin{tabular}{lccc}\\ \hline
      Sources of systematic uncertainty & $C_\mathrm{p,K}$ (in \%) & $C_\mathrm{Q,K}$ (in \%) & $C_\mathrm{Q,p}$ (in \%) \\ \hline
      Vertex $z$-position & 0.9--1.4  & $<$0.5 & $<$0.2 \\
      Centrality estimator & 1.3--6.9 & 0.1--0.8 & 0.3--1.3\\
      Pileup rejection & 0.1--1.4 & $<$0.3 & $<$0.2\\
      Space points in TPC & 0.6--3.8 & $<$0.5 & $<$0.2\\
      $\chi^{2}$ per space point in TPC & 0.7--1.2 & $<$0.4 & $<$0.2\\
      $\chi^{2}$ per space point in ITS & 0.3--2.8 & $<$0.4 & $<$0.2\\
      $\mathrm{DCA}_{xy}$ \& $\mathrm{DCA}_{z}$ & 1.6--3.9 & 0.4--1.2  & 0.3--1.0\\
      PID & 1.8--5.6 & 0.4--1.4 & 0.6--1.6\\ \hline
      Total & 5.5--8.8 & 0.8--2.1 & 1.1--1.8\\   
      \hline
      \end{tabular}
  \end{center}
\end{table}

The uncertainties in the measurements include both statistical and systematic components. A bootstrap approach~\cite{Luo:2014rea} is employed to evaluate statistical uncertainties. The systematic uncertainties are estimated by tightening and loosening the event and track selection criteria. The uncertainties related to the event selection stem from the position of the event vertex along the beam direction and the presence of pileup events. These contributions remain below 2\% across all centralities. To account for the systematic uncertainty in estimating the collision centrality, the centrality intervals are redefined using the midrapidity multiplicity distribution~\cite{ALICE:centrality1}. This modification introduces an uncertainty of up to 7\% in peripheral and less than 4\% in central collisions. The effects of the track selection are explored by varying the selection criteria on DCA, the number of reconstructed space points in the TPC, and the quality of the track fit, from their nominal values. Increasing the required number of TPC space points has a negligible effect, while variations in the DCA of the tracks in both transverse and longitudinal directions result in a systematic uncertainty of 1 to 3\%. The alteration of the $\chi^2$ per space point value in the TPC and the ITS leads to uncertainties of less than 2 and 4\%, respectively, across all centralities. Systematic uncertainties due to PID are assessed by varying the default selection criteria on $n(\sigma^\mathrm{TPC}_{i})$ and $n(\sigma^\mathrm{TPC+TOF}_{i})$ for each particle species. The default selection criterion is $2\sigma$, which is varied to 2.5$\sigma$, resulting in uncertainties ranging from 3\% to 6\%, depending on centrality. Finally, considering each source's contribution as uncorrelated, the total systematic uncertainty on the observables is determined by adding them in quadrature. Table~\ref{tab:sys} shows the summary of the contributions to the total systematic uncertainty on $C_\mathrm{p,K}$, $C_\mathrm{Q,K}$, and $C_\mathrm{Q,p}$.

\section{Results}

\label{sec:results}
\begin{figure}[ht]
  \begin{center}
    \includegraphics[width=0.65\linewidth]{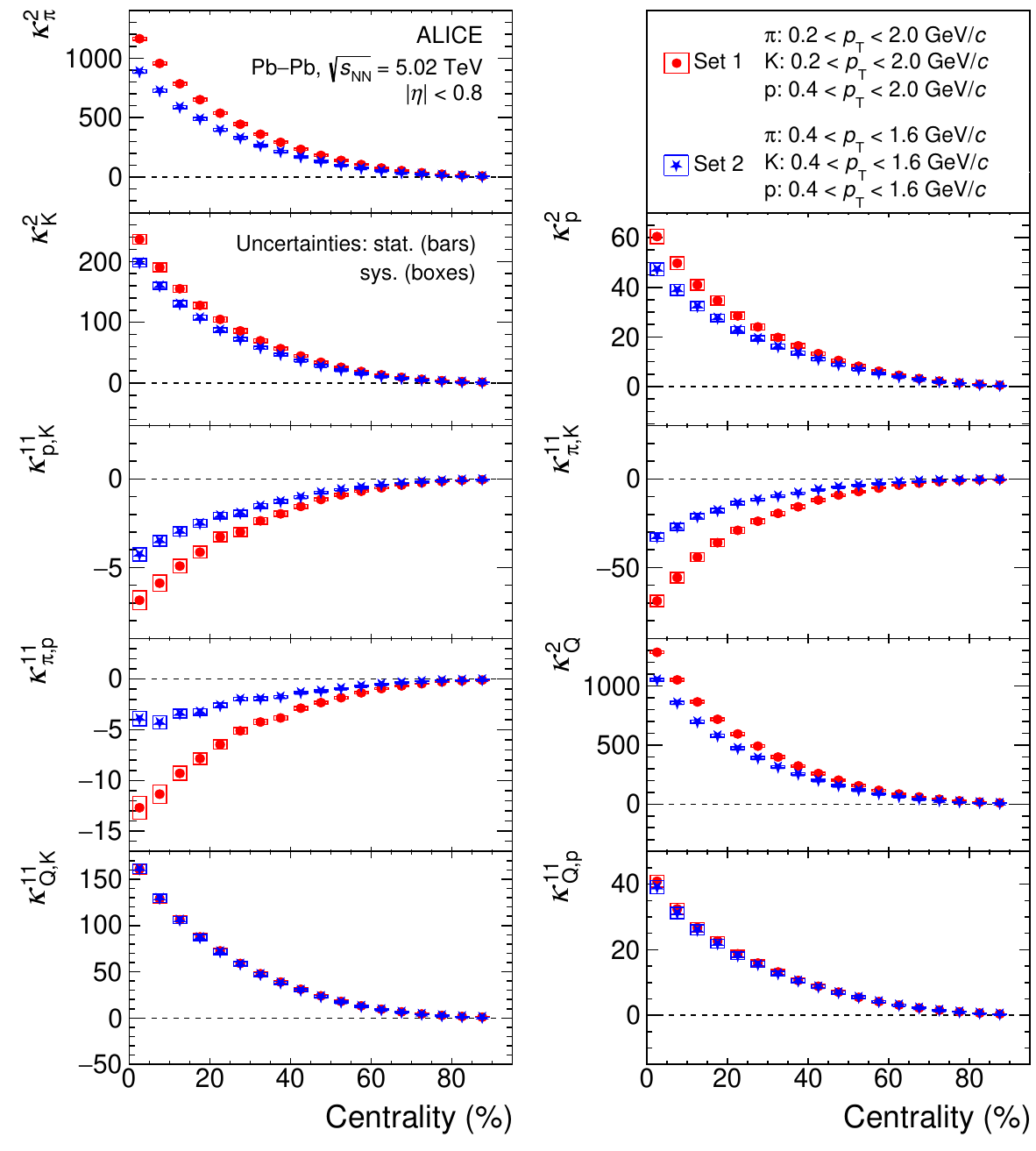}
  \end{center}
  \caption{Centrality dependence of diagonal cumulants ($\kappa^{2}_{\alpha}$) and off-diagonal cumulants ($\kappa^{11}_{\alpha,\beta}$) for the net-pion, net-kaon, net-proton, and net-charged particle distributions in Pb--Pb collisions at $\sqrt{s_\mathrm{NN}}=5.02$~TeV. The measurements are shown for two different $p_\mathrm{T}$ ranges. The statistical (systematic) uncertainties are represented by vertical bars (boxes).}
  \label{fig:CumAll}
\end{figure}

This analysis investigates the centrality dependence of the correlations $C_\mathrm{p,K}$, $C_\mathrm{Q,K}$, and $C_\mathrm{Q,p}$, for two different $p_\mathrm{T}$ ranges, referred to as Set 1 and Set 2. In case of Set 1, $\pi^{\pm}$ and $\mathrm{K}^{\pm}$ are selected in the range $0.2 < p_\mathrm{T} < 2.0$ GeV/$\it{c}$, while $\mathrm{p (\bar{p})}$ is selected within $0.4 < p_\mathrm{T} < 2.0$ GeV/$\it{c}$. In Set 2, $\mathrm{p (\bar{p})}$, $\pi^\pm$, and $\mathrm{K}^\pm$ are restricted to the narrower range $0.4 < p_\mathrm{T} < 1.6$ GeV/$\it{c}$. All measured diagonal and off-diagonal cumulants, $\kappa^2_\mathrm{\pi}$, $\kappa^2_\mathrm{K}$, $\kappa^2_\mathrm{p}$, $\kappa^{11}_\mathrm{p,K}$, $\kappa^{11}_\mathrm{\pi,K}$, $\kappa^{11}_\mathrm{\pi,p}$, $\kappa^2_\mathrm{Q}$, $\kappa^{11}_\mathrm{Q,K}$, and $\kappa^{11}_\mathrm{Q,p}$, are shown in Fig.~\ref{fig:CumAll} for both sets. The observed centrality dependence is consistent with the fact that cumulants are extensive quantities, scaling proportionally to the system volume. The diagonal cumulants $\kappa^2_\mathrm{\pi}$, $\kappa^2_\mathrm{K}$, $\kappa^2_\mathrm{p}$, and $\kappa^2_\mathrm{Q}$, along with the off-diagonal cumulants $\kappa_\mathrm{p,K}^{11}$, $\kappa_\mathrm{\pi,K}^{11}$, and $\kappa_\mathrm{\pi,p}^{11}$, show a dependence on the $p_\mathrm{T}$ interval. This behavior arises from the acceptance effect, where the chosen phase-space coverage influences the observed cumulant values~\cite{Ling:2015yau}. It should be noted that the corrections applied in this analysis address detector inefficiencies, such as tracking and PID, to recover the true particle numbers and their cumulants within the selected phase-space. However, no corrections are applied for kinematic acceptance (i.e., $p_\mathrm{T}$, $\eta$) itself, as this defines the physical region over which the measurement is done, consistent with previous studies~\cite{STAR:2020tga, ALICE:2022xpf}. The dependence of fluctuation observables on kinematic intervals is therefore a genuine physical effect, reflecting their sensitivity to the sampled region of the system and the underlying dynamics. On the other hand, $\kappa_\mathrm{Q,K}^{11}$ and $\kappa_\mathrm{Q,p}^{11}$ remain consistent within the uncertainties across the two $p_\mathrm{T}$ ranges, due to the interplay among their constituent terms: $\kappa^{11}_\mathrm{Q,K}=\kappa^{11}_\mathrm{p,K}+\kappa^{11}_\mathrm{\pi,K}+\kappa^{2}_\mathrm{K}$ and $\kappa^{11}_\mathrm{Q,p}=\kappa^{11}_\mathrm{p,K}+\kappa^{11}_\mathrm{\pi,p}+\kappa^{2}_\mathrm{p}$, leading to small variations. 

To explore the underlying physics mechanisms, the measurements are compared with model predictions from HIJING~\cite{Wang:1991hta}, EPOS LHC~\cite{Pierog:2013ria}, and Thermal-FIST (referred to as TheFIST in figures)~\cite{Vovchenko:2019pjl}, each offering a different approach to collision dynamics and hadronization. Resonance decays are included in all model calculations. The HIJING model (HIJING/$B\bar{B}$ v2.0~\cite{ToporPop:2004lve}) treats nucleus–nucleus collisions as a superposition of independent binary collisions of wounded nucleons, incorporating phenomena such as baryon junctions, mini-jet production, parton shadowing, and jet quenching, while excluding effects like thermal equilibrium and collectivity. The EPOS LHC model, on the other hand, introduces collective effects through effective parton ladder splitting, separating the high-density central core from the peripheral corona. This model, a parametrized version of EPOS 1.99~\cite{Pierog:2009zt}, does not fully utilize the comprehensive 3D hydrodynamic calculation and subsequent hadronic cascade employed in the original EPOS framework. The HRG model of Thermal-FIST, employs a thermal-statistical approach to hadronization, and determines hadron abundances at the freeze-out of inelastic interactions based on derivatives of the system's partition function. The key parameters of the model, the chemical freeze-out temperature ($T_\mathrm{chem}$), the volume corresponding to one unit of rapidity ($\mathrm{d}V/\mathrm{d}y$), and the strangeness saturation factor ($\gamma_{s}$)~\cite{Koch:1986ud, Rafelski:1991rh}, are tuned using the hadron yields measured by ALICE~\cite{ALICE:2019hno, ALICE:2014jbq, ALICE:2013cdo, ALICE:2013xmt} for different centralities. The $p_\mathrm{T}$ spectra are modeled using blast-wave fits from Ref.~\cite{ALICE:2019hno}. This model functions within both the grand-canonical ensemble (GCE) and canonical ensemble (CE) frameworks. In the GCE, the electric charge, baryon number, and strangeness are conserved only on average across the entire system, whereas in the CE, these quantities are conserved exactly within a fixed volume ($V_{c}$)~\cite{Becattini:2009sc, Vovchenko:2019kes}. The $V_{c}$, known as the correlation volume, is a defined region in phase space that ensures local conservation of QCD charges. In this approach, $V_{c} = k\mathrm{d}V/\mathrm{d}y$ implies that the system is truncated to $k$ units around midrapidity, i.e., $|y| < k/2$, making it equivalent to a global conservation model with a reduced rapidity cut-off~\cite{Vovchenko:2024pvk}. Therefore, in the $V_{c}$ approach, the size and geometry of the correlation volume impose certain limitations on fluctuation observables. For example, this approach provides accurate results for rapidities $|y_\mathrm{cut}|\lesssim k/4$ when fluctuations are measured within $|y| < y_\mathrm{cut}$, but fails at higher rapidities~\cite{Vovchenko:2024pvk}. There are also alternative methods for modeling local conservation of charges, based on the correlation length between particle and antiparticle in rapidity space~\cite{ALICE:2022xpf, Braun-Munzinger:2023gsd, Vovchenko:2024pvk}. In this analysis, the $V_{c}$ formulation is adopted, with $V_{c}$ initially fixed to $3\mathrm{d}V/\mathrm{d}y$ for the CE calculations, following the parametrization in Ref.~\cite{Vovchenko:2019kes}, which accurately describes the yield of light-flavored particles across all colliding systems with a precision better than 15\%. The recent ALICE measurements of the second-order cumulants of net-proton, net-$\Xi$, and the correlation between net-$\Xi$ and net-kaon are also described by the Thermal-FIST model with $V_{c} = 3\mathrm{d}V/\mathrm{d}y$, within experimental uncertainties~\cite{ALICE:2024rnr}.

An important distinction between the models considered in this work lies in their treatment of resonance decays. In HIJING, resonances are produced through string fragmentation and decay promptly according to fixed branching ratios, without further interaction in a hadronic medium~\cite{Wang:1991hta, ToporPop:2004lve}. As a result, the model does not capture medium-induced modifications to resonances, due to the absence of collective flow and hadronic rescattering. In contrast, EPOS LHC features a dynamically evolving treatment, in which resonances are produced both during the initial string fragmentation and at the hydrodynamic freeze-out surface~\cite{Pierog:2013ria, Knospe:2021jgt}. The resonances are allowed to decay according to their lifetimes and branching ratios without further medium-induced modifications. The Thermal-FIST model samples the resonances from a static thermal distribution at chemical freeze-out and performs a full decay chain of all unstable resonances based on branching ratios. The decay process is probabilistic in nature, meaning the actual number of decay products fluctuates from event-to-event~\cite{Vovchenko:2019pjl, Nahrgang:2014fza}. The decay kinematics are modeled assuming isotropic distributions in the resonance rest frame for two- and three-body channels, while many-body decays are treated approximately. No subsequent interactions such as rescattering or regeneration are considered post-decay.

\begin{figure}[t]
  \begin{center}
    \includegraphics[width=0.7\linewidth]{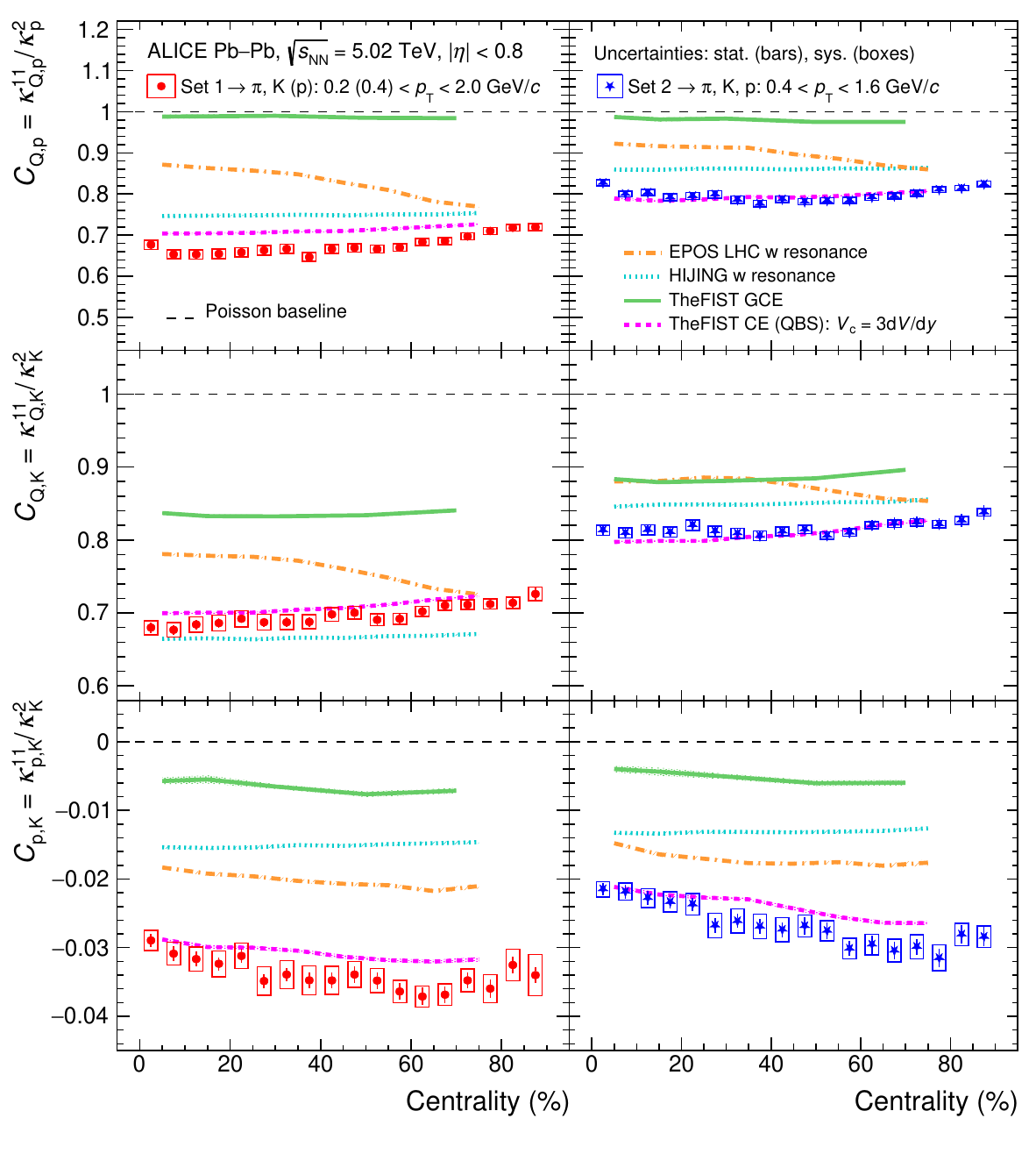}
  \end{center}
  \caption{Centrality dependence of correlations $C_\mathrm{Q,p}$ (top), $C_\mathrm{Q,K}$ (middle), and $C_\mathrm{p,K}$ (bottom) in Pb--Pb collisions at $\sqrt{s_\mathrm{NN}}=5.02$~TeV. The measurements are shown for two different $p_\mathrm{T}$ ranges in left and right columns. The predictions from HIJING~\cite{Wang:1991hta}, EPOS LHC~\cite{Pierog:2013ria}, and Thermal-FIST (TheFIST)~\cite{Vovchenko:2019pjl} model calculations with the grand canonical ensemble (GCE) and canonical ensemble (CE) formulation are denoted by colored lines. In the CE calculations, the electric charge (Q), baryon number (B), and strangeness (S) are conserved in a correlation volume of $V_c=3\mathrm{d}V/\mathrm{d}y$. The statistical (systematic) uncertainties are represented by vertical bars (boxes), and the dashed line corresponds to the Poisson baseline.}
  \label{fig:CQpQKpK}
\end{figure}

The correlations $C_\mathrm{Q,p}$, $C_\mathrm{Q,K}$, and $C_\mathrm{p,K}$ are shown in the top, middle, and bottom panels, respectively, in Fig.~\ref{fig:CQpQKpK}. The left column presents results obtained for Set 1 $p_\mathrm{T}$ acceptance, while the right column shows results for Set 2. All three correlations exhibit a weak centrality dependence, with $C_\mathrm{p,K}$ following a trend opposite to that of $C_\mathrm{Q,K}$ and $C_\mathrm{Q,p}$. This weak centrality dependence is expected, as these correlations are presented in terms of ratios, which are intensive quantities and thus largely insensitive to system size and volume fluctuations. The deviation from the Poisson baseline (unity for $C_\mathrm{Q,p}$ and $C_\mathrm{Q,K}$, and zero for $C_\mathrm{p,K}$) can be attributed to dynamic effects such as collectivity, resonance decays, and charge conservation~\cite{Garg:2013ata, Vovchenko:2020kwg, Braun-Munzinger:2019yxj, Bluhm:2020mpc, Mishra:2016qyj}. 

The HIJING model expectations are closer to the data for $C_\mathrm{Q,p}$ and $C_\mathrm{Q,K}$, but fail to capture the observed trends and magnitudes, with significant discrepancies in $C_\mathrm{p,K}$, mainly due to the incomplete modeling of resonance decays, which affect final-state interactions and the resulting particle correlations~\cite{Wang:1991hta, ToporPop:2004lve}. HIJING remains constant with centrality, possibly due to the absence of collective dynamics, whereas the EPOS LHC model shows a visible centrality dependence driven by varying contributions from the "core" and "corona" regions. EPOS LHC captures the decreasing trend of $C_\mathrm{p,K}$ with centrality, but predicts opposite trends for $C_\mathrm{Q,p}$ and $C_\mathrm{Q,K}$ compared to the data. The GCE results from Thermal-FIST model completely fail to describe the measurements in both Set 1 and Set 2 $p_\mathrm{T}$ acceptances. At the same time, the CE calculations show better agreement with the data compared to all the other models discussed. It is seen that CE captures both the magnitude and the centrality dependence of the correlations, particularly for the $p_\mathrm{T}$ acceptance in Set 2. This indicates that local conservation of charges in the CE formalism provides a more accurate explanation of the observed fluctuations and correlations in the experiment. However, further adjustments, such as varying $V_{c}$ in the CE formalism, may improve the model’s agreement with the data.

\begin{figure}[t]
  \begin{center}
    \includegraphics[width=\linewidth]{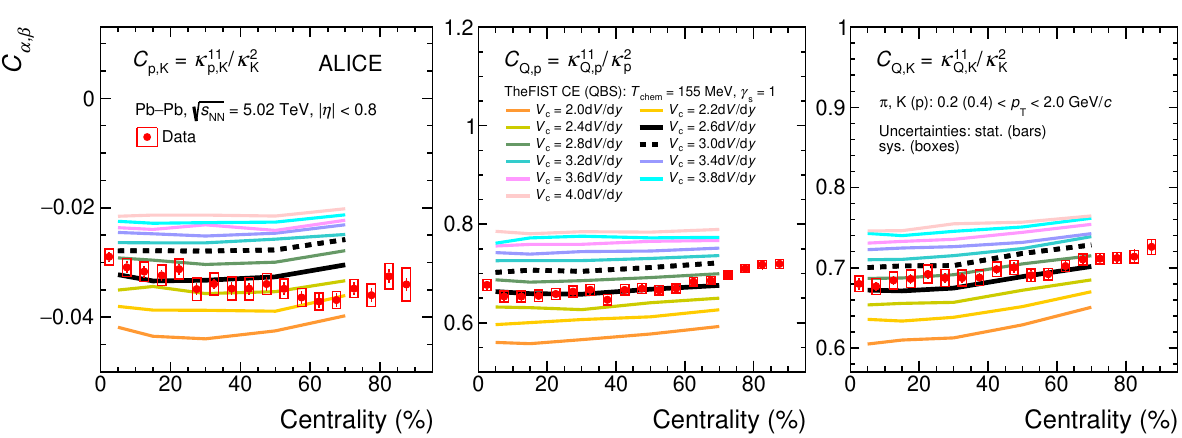}
  \end{center}
  \caption{(Color online) Correlations $C_\mathrm{p,K}$ (left), $C_\mathrm{Q,p}$ (middle), and $C_\mathrm{Q,K}$ (right) are shown as a function of centrality in Pb--Pb collisions at $\sqrt{s_\mathrm{NN}}=5.02$~TeV. Colored lines represent Thermal-FIST (TheFIST)~\cite{Vovchenko:2019pjl} model calculations for the canonical ensemble (CE) with different correlation volumes ($V_{c}$). The statistical (systematic) uncertainties are represented by vertical bars (boxes).}
  \label{fig:VcEffect}
\end{figure}

In Fig.~\ref{fig:VcEffect}, the correlations, $C_\mathrm{p,K}$, $C_\mathrm{Q,p}$, and $C_\mathrm{Q,K}$ for Set 1 $p_\mathrm{T}$ acceptance are compared with the Thermal-FIST results in the CE formalism, incorporating different $V_{c}$ values. The model parameters, $T_\mathrm{chem}$ and $\gamma_s$ are fixed to 155 MeV~\cite{Andronic:2017pug} and 1~\cite{Rafelski:1991rh}, respectively, across all centralities, while $V_{c}$ is systematically varied between 2$\mathrm{d}V/\mathrm{d}y$ and 4$\mathrm{d}V/\mathrm{d}y$. A quantitative comparison is performed by calculating the $\chi^2$ values between the experimental data and model predictions for varying $V_{c}$, considering only the statistical uncertainties. To account for all three correlations, a combined $\chi^2$ is defined as $\chi^2_\mathrm{combined}=\chi^{2}_{C_\mathrm{p,K}}+\chi^{2}_{C_\mathrm{Q,p}}+\chi^{2}_{C_\mathrm{Q,K}}$. The best estimate of $V_{c}$ is determined by minimizing $\chi^2_\mathrm{combined}$, with its statistical uncertainty calculated as half of the difference between the values of $V_{c}$ in which $\chi^2_\mathrm{combined}=\mathrm{MIN}(\chi^2_\mathrm{combined})+1$. Systematic uncertainties on $C_\mathrm{p,K}$, $C_\mathrm{Q,p}$, and $C_\mathrm{Q,K}$ are taken into account in the evaluation of $V_c$. The combined $\chi^2$ minimization is repeated after shifting the observables up and down by their respective systematic uncertainties, with half of the resulting difference assigned as the systematic uncertainty for $V_c$. The total uncertainty is then obtained by combining the statistical and systematic uncertainties in quadrature. This procedure yields $V_{c}=(2.60\pm 0.11)\mathrm{d}V/\mathrm{d}y$ for the Set 1 and $V_{c}=(2.82\pm 0.14)\mathrm{d}V/\mathrm{d}y$ for the Set 2 $p_\mathrm{T}$ acceptances. It is important to note that the extracted $V_{c}$ values may be affected by the treatment of resonance decays in the Thermal-FIST model, as resonances play a significant role in shaping particle yields and their correlations~\cite{Nahrgang:2014fza}. 

\begin{figure}[hb]
  \begin{center}
    \includegraphics[width=\linewidth]{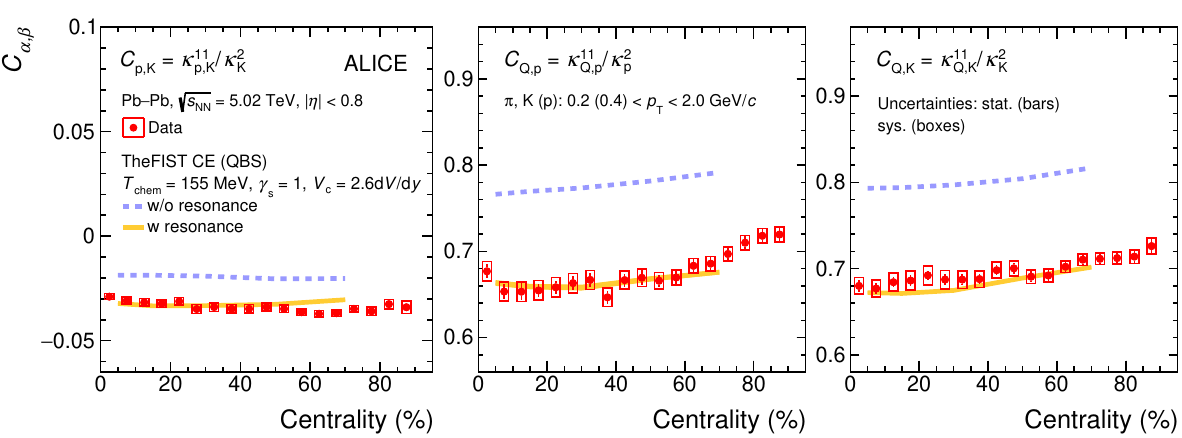}
  \end{center}
  \caption{Correlations $C_\mathrm{p,K}$ (left), $C_\mathrm{Q,p}$ (middle), and $C_\mathrm{Q,K}$ (right) are shown as a function of centrality in Pb--Pb collisions at $\sqrt{s_\mathrm{NN}}=5.02$~TeV. Colored lines represent Thermal-FIST (TheFIST)~\cite{Vovchenko:2019pjl} model calculations for canonical ensemble (CE) with (w) and without (w/o) resonance contribution. The statistical (systematic) uncertainties are represented by vertical bars (boxes).}
  \label{fig:Reso}
\end{figure}

Figure~\ref{fig:Reso} compares $C_\mathrm{p,K}$, $C_\mathrm{Q,p}$, and $C_\mathrm{Q,K}$ for Set 1 $p_\mathrm{T}$ acceptance with Thermal-FIST CE calculations (using model parameters $T_\mathrm{chem} = 155$ MeV, $\gamma_s = 1$, and $V_{c} = 2.6\mathrm{d}V/\mathrm{d}y$), both with and without the contribution of resonance decays. Including resonance decays in the model significantly enhances the magnitude of all three correlations across the full centrality range and allows the model to quantitatively reproduce the experimental data. In contrast, calculations without resonance decays fails to reproduce the observed results. The influence of resonances on the normalized second-order cumulants of net-pions and net-kaons has also been studied previously using the HIJING model in Ref.~\cite{ALICE:2022xpf}.

\begin{figure}[h]
  \begin{center}
    \includegraphics[width=0.8\linewidth]{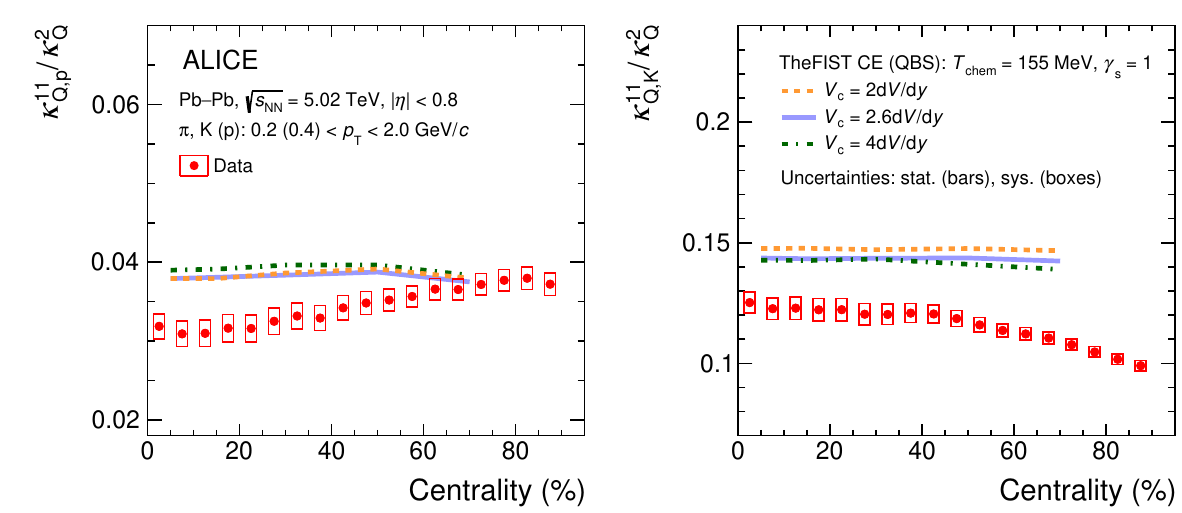}
  \end{center}
  \caption{Centrality dependence of $\kappa^{11}_\mathrm{Q,p}/\kappa^{2}_\mathrm{Q}$ (left) and $\kappa^{11}_\mathrm{Q,K}/\kappa^{2}_\mathrm{Q}$ (right) in Pb--Pb collisions at $\sqrt{s_\mathrm{NN}}=5.02$~TeV. Colored lines represent Thermal-FIST (TheFIST)~\cite{Vovchenko:2019pjl} model calculations for canonical ensemble (CE) with different correlation volumes ($V_{c}$). The statistical (systematic) uncertainties are represented by vertical bars (boxes).}
  \label{fig:NormQ}
\end{figure}

One can also study the ratios of the off-diagonal cumulants $\kappa^{11}_\mathrm{Q,p}$ and $\kappa^{11}_\mathrm{Q,K}$, to the diagonal cumulant $\kappa^{2}_\mathrm{Q}$ to eliminate the dependence on $V_{c}$. In these ratios, the $V_{c}$ dependence cancels out due to the same linear dependence on the acceptance fraction $\alpha_\mathrm{acc}$ of both the numerator and the denominator~\cite{Vovchenko:2020gne}. The measurements of $\kappa^{11}_\mathrm{Q,p}/\kappa^{2}_\mathrm{Q}$ and $\kappa^{11}_\mathrm{Q,K}/\kappa^{2}_\mathrm{Q}$ for Set 1 $p_\mathrm{T}$ interval, compared with model results, are shown in the left and right panels of Fig.~\ref{fig:NormQ}, respectively. The Thermal-FIST model results with different $V_{c}$ values are comparable despite small deviations, which can be attributed to the imperfect cancellation of the acceptance factor (the net-charged particle is defined as the sum of the net-pion, net-kaon, and net-proton) and the specific implementation of $V_{c}$ in the model~\cite{Vovchenko:2024pvk}. The model could not fully capture the centrality dependence of the data, and overestimates its magnitude. This discrepancy could be due to an incomplete description of the resonance decays, assumptions in the $V_c$ approach, or missing effects such as initial-state fluctuations and final-state interactions.

\begin{figure}[h]
  \begin{center}
    \includegraphics[width=0.8\linewidth]{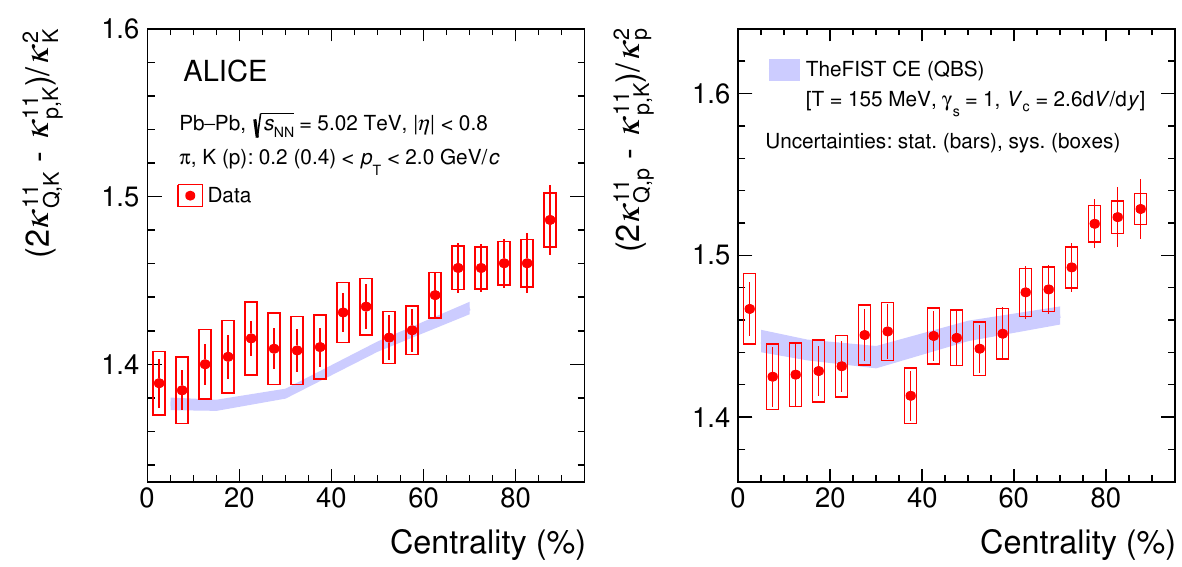}
  \end{center}
  \caption{Centrality dependence of $(2\kappa^{11}_\mathrm{Q,K}-\kappa^{11}_\mathrm{p,K})/\kappa^{2}_\mathrm{K}$ (left) and $(2\kappa^{11}_\mathrm{Q,p}-\kappa^{11}_\mathrm{p,K})/\kappa^{2}_\mathrm{p}$ (right) in Pb--Pb collisions at $\sqrt{s_\mathrm{NN}}=5.02$~TeV. Thermal-FIST (TheFIST)~\cite{Vovchenko:2019pjl} model calculation for canonical ensemble (CE) with a correlation volume of $V_\mathrm{c}=2.6\mathrm{d}V/\mathrm{d}y$ is represented by the colored band. The statistical (systematic) uncertainties are represented by vertical bars (boxes).}
  \label{fig:Mag}
\end{figure} 

\begin{figure}[!htbp]
  \begin{center}
    \includegraphics[width=0.8\linewidth]{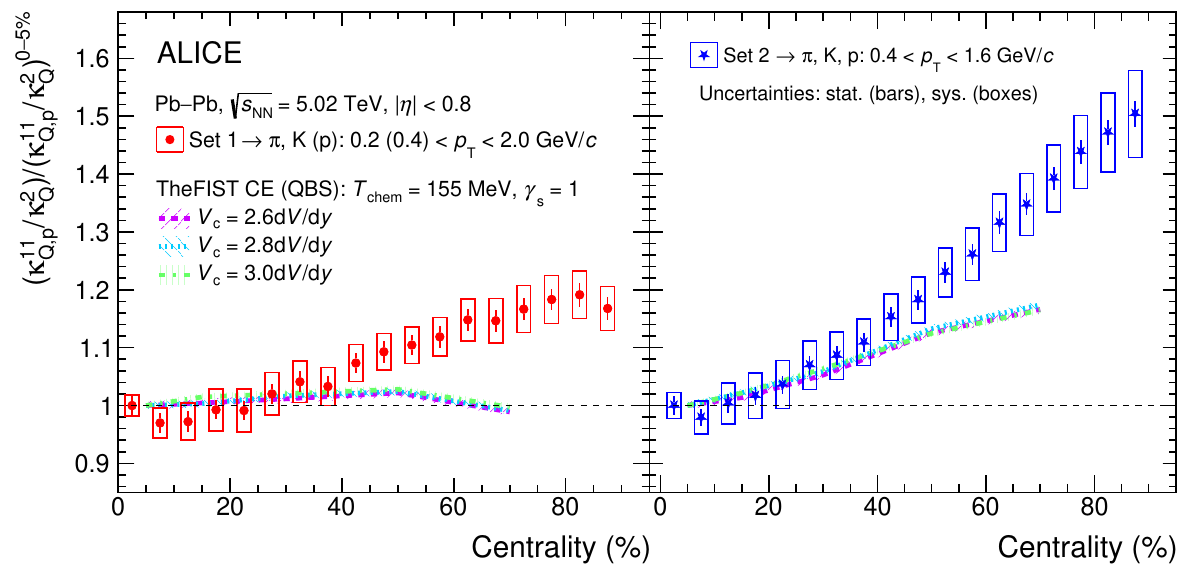}
  \end{center}
  \caption{Centrality dependence of $(\kappa^{11}_\mathrm{Q,p}/\kappa^{2}_\mathrm{Q})/(\kappa^{11}_\mathrm{Q,p}/\kappa^{2}_\mathrm{Q})^{0-5\%}$ in Pb--Pb collisions at $\sqrt{s_\mathrm{NN}}=5.02$~TeV for Set 1 (left) and Set 2 (right) $p_\mathrm{T}$ acceptances. The colored bands represent the Thermal-FIST (TheFIST)~\cite{Vovchenko:2019pjl} model calculations for canonical ensemble (CE) with different $V_{c}$ values. The statistical (systematic) uncertainties are represented by vertical bars (boxes).}
  \label{fig:Mag2}
\end{figure}

Another possible factor affecting the cumulants and their ratios is the initial magnetic field generated by the spectator protons~\cite{Ding:2021cwv, Ding:2023bft}. Figure~\ref{fig:Mag} shows the centrality dependence of $(2\kappa^{11}_\mathrm{Q,K}-\kappa^{11}_\mathrm{p,K})/\kappa^{2}_\mathrm{K}$ and $(2\kappa^{11}_\mathrm{Q,p}-\kappa^{11}_\mathrm{p,K})/\kappa^{2}_\mathrm{p}$, which have been proposed as observables sensitive to the magnetic field~\cite{Ding:2021cwv}. Both measurements exhibit a subtle increase from semi-central to peripheral collisions. The ratio $(2\kappa^{11}_\mathrm{Q,K}-\kappa^{11}_\mathrm{p,K})/\kappa^{2}_\mathrm{K}$ rises by $\sim 4\%$ from the 50--55\% to 85--90\% centrality class, while $(2\kappa^{11}_\mathrm{Q,p}-\kappa^{11}_\mathrm{p,K})/\kappa^{2}_\mathrm{p}$ increases by $\sim 5\%$, with relative significances of 2.1$\sigma$ and 2.8$\sigma$, respectively. Despite the absence of the magnetic field effect in the model, the Thermal-FIST CE calculations describe the measurements within experimental uncertainties. While these ratios do not show a clear signature of the magnetic field, the scaled ratio $(\kappa^{11}_\mathrm{Q,p}/\kappa^{2}_\mathrm{Q})/(\kappa^{11}_\mathrm{Q,p}/\kappa^{2}_\mathrm{Q})^{0-5\%}$ exhibits a more pronounced centrality dependence, as shown in Fig.~\ref{fig:Mag2}. For both Set 1 (left) and Set 2 (right), this ratio increases with centrality, following a trend consistent with LQCD predictions~\cite{Ding:2023bft}. In Set 1, the observable deviates from unity beyond 40\% centrality, increasing by nearly 20\% at 85--90\% centrality, while in Set 2, the deviation starts at 25\% centrality and reaches almost 50\% at 85--90\% centrality. The measurements are compared to Thermal-FIST CE calculations without magnetic field for $V_{c}$ values of $2.6\mathrm{d}V/\mathrm{d}y$, $2.8\mathrm{d}V/\mathrm{d}y$ and $3.0\mathrm{d}V/\mathrm{d}y$, with
model results consistent across these $V_c$ values. The model fails to explain the deviation beyond 50\% centrality, which could result from incomplete resonance decay descriptions, missing physical processes, or possibly an influence from the initial magnetic field. Further investigation is required to confirm whether a magnetic field effect contributes to these trends or if other factors dominate.

\begin{figure}[!htbp]
  \begin{center}
    \includegraphics[width=\linewidth]{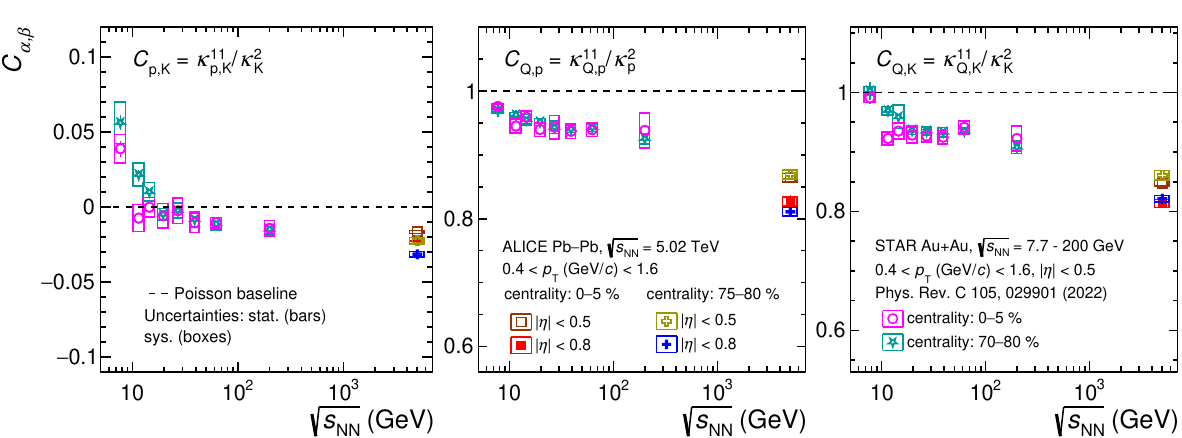}
  \end{center}
  \caption{Energy dependence of $C_\mathrm{p,K}$ (left), $C_\mathrm{Q,p}$ (middle), and $C_\mathrm{Q,K}$ (right), with lower energy results corresponding to STAR experiment~\cite{STAR:2019ans}. The statistical (systematic) uncertainties are represented by vertical bars (boxes) and the dashed line corresponds to Poisson baseline.}
  \label{fig:Energy}
\end{figure}

Figure~\ref{fig:Energy} shows $C_\mathrm{p,K}$, $C_\mathrm{Q,p}$, and $C_\mathrm{Q,K}$ as a function of collision energy for central and peripheral events, combining ALICE data at $\sqrt{s_\mathrm{NN}}=5.02$~TeV for Pb--Pb collisions and STAR data for Au+Au collisions from $\sqrt{s_\mathrm{NN}}=7.7-200$ GeV~\cite{STAR:2019ans}. While the STAR results correspond to $|\eta| < 0.5$, the ALICE measurements are presented for both $|\eta| < 0.5$ and $|\eta| < 0.8$, offering insights into the effect of increasing acceptance window. All three correlations exhibit a monotonic decrease from RHIC to LHC energies. Notably, the deviation from the Poisson baseline increases with energy for both central and peripheral events, which is due to several differences in the
underlying physics processes. A key factor is the decreasing beam rapidity with decreasing energy, which leads to a higher number of stopped protons in the kinematic acceptance and an increase in the $\alpha_\mathrm{acc}$ \textemdash{} both crucial for the description of global and local charge conservation. In addition, volume fluctuations, which cannot be completely suppressed even in narrow centrality bins, may contribute to the observed trend.

\section{Summary}
\label{sec:summary}
This paper presents the centrality dependence of correlations among the net-charged particle, net-proton, and net-kaon ($C_\mathrm{p,K}$, $C_\mathrm{Q,p}$, and $C_\mathrm{Q,K}$) measured in Pb--Pb collisions at $\sqrt{s_\mathrm{NN}}=5.02$~TeV, with net-proton and net-kaon serving as proxies for the net-baryon and the net-strangeness. All three correlations show pronounced deviations from the Poisson baseline, with resonance decays and charge conservation being significant contributing effects. Model calculations from HIJING, EPOS LHC, and Thermal-FIST in the grand canonical ensemble fail to explain the measurements. However, the Thermal-FIST model in the canonical ensemble, based on local charge conservation of the electric charge, baryon number, and strangeness, exhibits a better agreement with the data across all centralities. A combined $\chi^{2}$ analysis of the correlations yields estimates of the correlation volume as $(2.60 \pm 0.11)\mathrm{d}V/\mathrm{d}y$ for the $p_\mathrm{T}$ range $0.2 < p_\mathrm{T} < 2.0$ GeV/$\it{c}$, and $(2.82 \pm 0.14)\mathrm{d}V/\mathrm{d}y$ for $0.4 < p_\mathrm{T} < 1.6$ GeV/$\it{c}$. The obtained correlation volumes \textemdash{} a specific implementation of the Thermal-FIST model \textemdash{} may be influenced by an incomplete implementation of resonance decays, assumptions in the $V_c$ approach, or missing dynamical effects such as initial-state fluctuations and final-state interactions. The ratios $\kappa^{11}_\mathrm{Q,p}/\kappa^{2}_\mathrm{Q}$ and $\kappa^{11}_\mathrm{Q,K}/\kappa^{2}_\mathrm{Q}$, which are independent from the choice of $V_{c}$, reveal a notable deviation from the Thermal-FIST model calculations, which underscores the need for a better understanding of the underlying physics processes. Lattice QCD calculations predict that specific combinations of susceptibilities for the electric charge, baryon number, and strangeness are sensitive to the magnetic field, and these are studied experimentally using their corresponding proxies. While $(2\kappa^{11}_\mathrm{Q,K}-\kappa^{11}_\mathrm{p,K})/\kappa^{2}_\mathrm{K}$ and $(2\kappa^{11}_\mathrm{Q,p}-\kappa^{11}_\mathrm{p,K})/\kappa^{2}_\mathrm{p}$ display a subtle increase from semicentral to peripheral collisions, the scaled ratio $(\kappa^{11}_\mathrm{Q,p}/\kappa^{2}_\mathrm{Q})/(\kappa^{11}_\mathrm{Q,p}/\kappa^{2}_\mathrm{Q})^{0-5\%}$ shows a more pronounced increase, in qualitative agreement with LQCD predictions. Thermal-FIST calculations without magnetic field fail to reproduce the observed trend beyond 50\% centrality, indicating the need for further investigations to confirm the magnetic field effect and rule out other possible explanations. The collision energy dependence of $C_\mathrm{p,K}$, $C_\mathrm{Q,p}$, and $C_\mathrm{Q,K}$ shows a decrease from RHIC to LHC energies, with larger deviations from the Poisson baseline at higher energies. Understanding the impact of physics mechanisms such as stopped protons and changes in beam rapidity, is crucial for accurately modeling global and local charge conservation and correcting for volume fluctuations.

%%%%%%%%%%%%%%%%%%%%%%%%%%%%%%%%
 % end main text 
%%%%%%%%%%%%%%%%%%%%%%%%%%%%%%%%

%%%%% acknowledgements - handled by EB chairs 
\newenvironment{acknowledgement}{\relax}{\relax}
\begin{acknowledgement}
\section*{Acknowledgements}
% add specific acknowledgements here 
% ...but please don't remove the line below: funding agencies
% will be acknowledged with a custom tex file handled by EB chairs after Collab Round 2
% Version: 2025-03-06

The ALICE Collaboration would like to thank all its engineers and technicians for their invaluable contributions to the construction of the experiment and the CERN accelerator teams for the outstanding performance of the LHC complex.
The ALICE Collaboration gratefully acknowledges the resources and support provided by all Grid centres and the Worldwide LHC Computing Grid (WLCG) collaboration.
The ALICE Collaboration acknowledges the following funding agencies for their support in building and running the ALICE detector:
A. I. Alikhanyan National Science Laboratory (Yerevan Physics Institute) Foundation (ANSL), State Committee of Science and World Federation of Scientists (WFS), Armenia;
Austrian Academy of Sciences, Austrian Science Fund (FWF): [M 2467-N36] and Nationalstiftung f\"{u}r Forschung, Technologie und Entwicklung, Austria;
Ministry of Communications and High Technologies, National Nuclear Research Center, Azerbaijan;
Conselho Nacional de Desenvolvimento Cient\'{\i}fico e Tecnol\'{o}gico (CNPq), Financiadora de Estudos e Projetos (Finep), Funda\c{c}\~{a}o de Amparo \`{a} Pesquisa do Estado de S\~{a}o Paulo (FAPESP) and Universidade Federal do Rio Grande do Sul (UFRGS), Brazil;
Bulgarian Ministry of Education and Science, within the National Roadmap for Research Infrastructures 2020-2027 (object CERN), Bulgaria;
Ministry of Education of China (MOEC) , Ministry of Science \& Technology of China (MSTC) and National Natural Science Foundation of China (NSFC), China;
Ministry of Science and Education and Croatian Science Foundation, Croatia;
Centro de Aplicaciones Tecnol\'{o}gicas y Desarrollo Nuclear (CEADEN), Cubaenerg\'{\i}a, Cuba;
Ministry of Education, Youth and Sports of the Czech Republic, Czech Republic;
The Danish Council for Independent Research | Natural Sciences, the VILLUM FONDEN and Danish National Research Foundation (DNRF), Denmark;
Helsinki Institute of Physics (HIP), Finland;
Commissariat \`{a} l'Energie Atomique (CEA) and Institut National de Physique Nucl\'{e}aire et de Physique des Particules (IN2P3) and Centre National de la Recherche Scientifique (CNRS), France;
Bundesministerium f\"{u}r Bildung und Forschung (BMBF) and GSI Helmholtzzentrum f\"{u}r Schwerionenforschung GmbH, Germany;
General Secretariat for Research and Technology, Ministry of Education, Research and Religions, Greece;
National Research, Development and Innovation Office, Hungary;
Department of Atomic Energy Government of India (DAE), Department of Science and Technology, Government of India (DST), University Grants Commission, Government of India (UGC) and Council of Scientific and Industrial Research (CSIR), India;
National Research and Innovation Agency - BRIN, Indonesia;
Istituto Nazionale di Fisica Nucleare (INFN), Italy;
Japanese Ministry of Education, Culture, Sports, Science and Technology (MEXT) and Japan Society for the Promotion of Science (JSPS) KAKENHI, Japan;
Consejo Nacional de Ciencia (CONACYT) y Tecnolog\'{i}a, through Fondo de Cooperaci\'{o}n Internacional en Ciencia y Tecnolog\'{i}a (FONCICYT) and Direcci\'{o}n General de Asuntos del Personal Academico (DGAPA), Mexico;
Nederlandse Organisatie voor Wetenschappelijk Onderzoek (NWO), Netherlands;
The Research Council of Norway, Norway;
Pontificia Universidad Cat\'{o}lica del Per\'{u}, Peru;
Ministry of Science and Higher Education, National Science Centre and WUT ID-UB, Poland;
Korea Institute of Science and Technology Information and National Research Foundation of Korea (NRF), Republic of Korea;
Ministry of Education and Scientific Research, Institute of Atomic Physics, Ministry of Research and Innovation and Institute of Atomic Physics and Universitatea Nationala de Stiinta si Tehnologie Politehnica Bucuresti, Romania;
Ministerstvo skolstva, vyskumu, vyvoja a mladeze SR, Slovakia;
National Research Foundation of South Africa, South Africa;
Swedish Research Council (VR) and Knut \& Alice Wallenberg Foundation (KAW), Sweden;
European Organization for Nuclear Research, Switzerland;
Suranaree University of Technology (SUT), National Science and Technology Development Agency (NSTDA) and National Science, Research and Innovation Fund (NSRF via PMU-B B05F650021), Thailand;
Turkish Energy, Nuclear and Mineral Research Agency (TENMAK), Turkey;
National Academy of  Sciences of Ukraine, Ukraine;
Science and Technology Facilities Council (STFC), United Kingdom;
National Science Foundation of the United States of America (NSF) and United States Department of Energy, Office of Nuclear Physics (DOE NP), United States of America.
In addition, individual groups or members have received support from:
Czech Science Foundation (grant no. 23-07499S), Czech Republic;
FORTE project, reg.\ no.\ CZ.02.01.01/00/22\_008/0004632, Czech Republic, co-funded by the European Union, Czech Republic;
European Research Council (grant no. 950692), European Union;
Deutsche Forschungs Gemeinschaft (DFG, German Research Foundation) ``Neutrinos and Dark Matter in Astro- and Particle Physics'' (grant no. SFB 1258), Germany;
ICSC - National Research Center for High Performance Computing, Big Data and Quantum Computing and FAIR - Future Artificial Intelligence Research, funded by the NextGenerationEU program (Italy).

\end{acknowledgement}

%%%%%%%% Bibliography
\bibliographystyle{utphys}   % Remember we use title in the biblio
\bibliography{bibliography2}
%\input {bibliography.tex}  

%%%%%%%%%%%%%%%%%%%%%%%%%%%%%%%%
% Appendices: yours (if any) + authorlist
%%%%%%%%%%%%%%%%%%%%%%%%%%%%%%%%
\newpage
\appendix

%
%\input{} % put your appendices here (if any)
%

%%%%% Authorlist - please do not touch: handled by EB chairs 
\section{The ALICE Collaboration}
\label{app:collab}
% ALICE Collaboration author list for 2025-03-06
\begin{flushleft} 
\small

S.~Acharya\,\orcidlink{0000-0002-9213-5329}\,$^{\rm 50}$, 
A.~Agarwal$^{\rm 133}$, 
G.~Aglieri Rinella\,\orcidlink{0000-0002-9611-3696}\,$^{\rm 32}$, 
L.~Aglietta\,\orcidlink{0009-0003-0763-6802}\,$^{\rm 24}$, 
M.~Agnello\,\orcidlink{0000-0002-0760-5075}\,$^{\rm 29}$, 
N.~Agrawal\,\orcidlink{0000-0003-0348-9836}\,$^{\rm 25}$, 
Z.~Ahammed\,\orcidlink{0000-0001-5241-7412}\,$^{\rm 133}$, 
S.~Ahmad\,\orcidlink{0000-0003-0497-5705}\,$^{\rm 15}$, 
S.U.~Ahn\,\orcidlink{0000-0001-8847-489X}\,$^{\rm 71}$, 
I.~Ahuja\,\orcidlink{0000-0002-4417-1392}\,$^{\rm 36}$
% ...

\section*{Affiliation Notes}

$^{\rm I}$ Deceased\\
$^{\rm II}$ Also at: Max-Planck-Institut fur Physik, Munich, Germany\\
$^{\rm III}$ Also at: Italian National Agency for New Technologies, Energy and Sustainable Economic Development (ENEA), Bologna, Italy\\
$^{\rm IV}$ Also at: Instituto de Fisica da Universidade de Sao Paulo\\
$^{\rm V}$ Also at: Dipartimento DET del Politecnico di Torino, Turin, Italy\\
$^{\rm VI}$ Also at: Department of Applied Physics, Aligarh Muslim University, Aligarh, India\\
$^{\rm VII}$ Also at: Institute of Theoretical Physics, University of Wroclaw, Poland\\
$^{\rm VIII}$ Also at: Facultad de Ciencias, Universidad Nacional Aut\'{o}noma de M\'{e}xico, Mexico City, Mexico\\

\section*{Collaboration Institutes}

$^{1}$ A.I. Alikhanyan National Science Laboratory (Yerevan Physics Institute) Foundation, Yerevan, Armenia\\
$^{2}$ AGH University of Krakow, Cracow, Poland\\
$^{3}$ Bogolyubov Institute for Theoretical Physics, National Academy of Sciences of Ukraine, Kiev, Ukraine\\
$^{4}$ Bose Institute, Department of Physics  and Centre for Astroparticle Physics and Space Science (CAPSS), Kolkata, India\\
$^{5}$ California Polytechnic State University, San Luis Obispo, California, United States\\
$^{6}$ Central China Normal University, Wuhan, China\\
$^{7}$ Centro de Aplicaciones Tecnol\'{o}gicas y Desarrollo Nuclear (CEADEN), Havana, Cuba\\
$^{8}$ Centro de Investigaci\'{o}n y de Estudios Avanzados (CINVESTAV), Mexico City and M\'{e}rida, Mexico\\
$^{9}$ Chicago State University, Chicago, Illinois, United States\\
$^{10}$ China Nuclear Data Center, China Institute of Atomic Energy, Beijing, China\\
$^{11}$ China University of Geosciences, Wuhan, China\\
$^{12}$ Chungbuk National University, Cheongju, Republic of Korea\\
$^{13}$ Comenius University Bratislava, Faculty of Mathematics, Physics and Informatics, Bratislava, Slovak Republic\\
$^{14}$ Creighton University, Omaha, Nebraska, United States\\
$^{15}$ Department of Physics, Aligarh Muslim University, Aligarh, India\\
$^{16}$ Department of Physics, Pusan National University, Pusan, Republic of Korea\\
$^{17}$ Department of Physics, Sejong University, Seoul, Republic of Korea\\
$^{18}$ Department of Physics, University of California, Berkeley, California, United States\\
$^{19}$ Department of Physics, University of Oslo, Oslo, Norway\\
$^{20}$ Department of Physics and Technology, University of Bergen, Bergen, Norway\\
$^{21}$ Dipartimento di Fisica, Universit\`{a} di Pavia, Pavia, Italy\\
$^{22}$ Dipartimento di Fisica dell'Universit\`{a} and Sezione INFN, Cagliari, Italy\\
$^{23}$ Dipartimento di Fisica dell'Universit\`{a} and Sezione INFN, Trieste, Italy\\
$^{24}$ Dipartimento di Fisica dell'Universit\`{a} and Sezione INFN, Turin, Italy\\
$^{25}$ Dipartimento di Fisica e Astronomia dell'Universit\`{a} and Sezione INFN, Bologna, Italy\\
$^{26}$ Dipartimento di Fisica e Astronomia dell'Universit\`{a} and Sezione INFN, Catania, Italy\\
$^{27}$ Dipartimento di Fisica e Astronomia dell'Universit\`{a} and Sezione INFN, Padova, Italy\\
$^{28}$ Dipartimento di Fisica `E.R.~Caianiello' dell'Universit\`{a} and Gruppo Collegato INFN, Salerno, Italy\\
$^{29}$ Dipartimento DISAT del Politecnico and Sezione INFN, Turin, Italy\\
$^{30}$ Dipartimento di Scienze MIFT, Universit\`{a} di Messina, Messina, Italy\\
$^{31}$ Dipartimento Interateneo di Fisica `M.~Merlin' and Sezione INFN, Bari, Italy\\
$^{32}$ European Organization for Nuclear Research (CERN), Geneva, Switzerland\\
$^{33}$ Faculty of Electrical Engineering, Mechanical Engineering and Naval Architecture, University of Split, Split, Croatia\\
$^{34}$ Faculty of Nuclear Sciences and Physical Engineering, Czech Technical University in Prague, Prague, Czech Republic\\
$^{35}$ Faculty of Physics, Sofia University, Sofia, Bulgaria\\
$^{36}$ Faculty of Science, P.J.~\v{S}af\'{a}rik University, Ko\v{s}ice, Slovak Republic\\
$^{37}$ Faculty of Technology, Environmental and Social Sciences, Bergen, Norway\\
$^{38}$ Frankfurt Institute for Advanced Studies, Johann Wolfgang Goethe-Universit\"{a}t Frankfurt, Frankfurt, Germany\\
$^{39}$ Fudan University, Shanghai, China\\
$^{40}$ Gangneung-Wonju National University, Gangneung, Republic of Korea\\
$^{41}$ Gauhati University, Department of Physics, Guwahati, India\\
$^{42}$ Helmholtz-Institut f\"{u}r Strahlen- und Kernphysik, Rheinische Friedrich-Wilhelms-Universit\"{a}t Bonn, Bonn, Germany\\
$^{43}$ Helsinki Institute of Physics (HIP), Helsinki, Finland\\
$^{44}$ High Energy Physics Group,  Universidad Aut\'{o}noma de Puebla, Puebla, Mexico\\
$^{45}$ Horia Hulubei National Institute of Physics and Nuclear Engineering, Bucharest, Romania\\
$^{46}$ HUN-REN Wigner Research Centre for Physics, Budapest, Hungary\\
$^{47}$ Indian Institute of Technology Bombay (IIT), Mumbai, India\\
$^{48}$ Indian Institute of Technology Indore, Indore, India\\
$^{49}$ INFN, Laboratori Nazionali di Frascati, Frascati, Italy\\
$^{50}$ INFN, Sezione di Bari, Bari, Italy\\
$^{51}$ INFN, Sezione di Bologna, Bologna, Italy\\
$^{52}$ INFN, Sezione di Cagliari, Cagliari, Italy\\
$^{53}$ INFN, Sezione di Catania, Catania, Italy\\
$^{54}$ INFN, Sezione di Padova, Padova, Italy\\
$^{55}$ INFN, Sezione di Pavia, Pavia, Italy\\
$^{56}$ INFN, Sezione di Torino, Turin, Italy\\
$^{57}$ INFN, Sezione di Trieste, Trieste, Italy\\
$^{58}$ Inha University, Incheon, Republic of Korea\\
$^{59}$ Institute for Gravitational and Subatomic Physics (GRASP), Utrecht University/Nikhef, Utrecht, Netherlands\\
$^{60}$ Institute of Experimental Physics, Slovak Academy of Sciences, Ko\v{s}ice, Slovak Republic\\
$^{61}$ Institute of Physics, Homi Bhabha National Institute, Bhubaneswar, India\\
$^{62}$ Institute of Physics of the Czech Academy of Sciences, Prague, Czech Republic\\
$^{63}$ Institute of Space Science (ISS), Bucharest, Romania\\
$^{64}$ Institut f\"{u}r Kernphysik, Johann Wolfgang Goethe-Universit\"{a}t Frankfurt, Frankfurt, Germany\\
$^{65}$ Instituto de Ciencias Nucleares, Universidad Nacional Aut\'{o}noma de M\'{e}xico, Mexico City, Mexico\\
$^{66}$ Instituto de F\'{i}sica, Universidade Federal do Rio Grande do Sul (UFRGS), Porto Alegre, Brazil\\
$^{67}$ Instituto de F\'{\i}sica, Universidad Nacional Aut\'{o}noma de M\'{e}xico, Mexico City, Mexico\\
$^{68}$ iThemba LABS, National Research Foundation, Somerset West, South Africa\\
$^{69}$ Jeonbuk National University, Jeonju, Republic of Korea\\
$^{70}$ Johann-Wolfgang-Goethe Universit\"{a}t Frankfurt Institut f\"{u}r Informatik, Fachbereich Informatik und Mathematik, Frankfurt, Germany\\
$^{71}$ Korea Institute of Science and Technology Information, Daejeon, Republic of Korea\\
$^{72}$ Laboratoire de Physique Subatomique et de Cosmologie, Universit\'{e} Grenoble-Alpes, CNRS-IN2P3, Grenoble, France\\
$^{73}$ Lawrence Berkeley National Laboratory, Berkeley, California, United States\\
$^{74}$ Lund University Department of Physics, Division of Particle Physics, Lund, Sweden\\
$^{75}$ Nagasaki Institute of Applied Science, Nagasaki, Japan\\
$^{76}$ Nara Women{'}s University (NWU), Nara, Japan\\
$^{77}$ National and Kapodistrian University of Athens, School of Science, Department of Physics , Athens, Greece\\
$^{78}$ National Centre for Nuclear Research, Warsaw, Poland\\
$^{79}$ National Institute of Science Education and Research, Homi Bhabha National Institute, Jatni, India\\
$^{80}$ National Nuclear Research Center, Baku, Azerbaijan\\
$^{81}$ National Research and Innovation Agency - BRIN, Jakarta, Indonesia\\
$^{82}$ Niels Bohr Institute, University of Copenhagen, Copenhagen, Denmark\\
$^{83}$ Nikhef, National institute for subatomic physics, Amsterdam, Netherlands\\
$^{84}$ Nuclear Physics Group, STFC Daresbury Laboratory, Daresbury, United Kingdom\\
$^{85}$ Nuclear Physics Institute of the Czech Academy of Sciences, Husinec-\v{R}e\v{z}, Czech Republic\\
$^{86}$ Oak Ridge National Laboratory, Oak Ridge, Tennessee, United States\\
$^{87}$ Ohio State University, Columbus, Ohio, United States\\
$^{88}$ Physics department, Faculty of science, University of Zagreb, Zagreb, Croatia\\
$^{89}$ Physics Department, Panjab University, Chandigarh, India\\
$^{90}$ Physics Department, University of Jammu, Jammu, India\\
$^{91}$ Physics Program and International Institute for Sustainability with Knotted Chiral Meta Matter (WPI-SKCM$^{2}$), Hiroshima University, Hiroshima, Japan\\
$^{92}$ Physikalisches Institut, Eberhard-Karls-Universit\"{a}t T\"{u}bingen, T\"{u}bingen, Germany\\
$^{93}$ Physikalisches Institut, Ruprecht-Karls-Universit\"{a}t Heidelberg, Heidelberg, Germany\\
$^{94}$ Physik Department, Technische Universit\"{a}t M\"{u}nchen, Munich, Germany\\
$^{95}$ Politecnico di Bari and Sezione INFN, Bari, Italy\\
$^{96}$ Research Division and ExtreMe Matter Institute EMMI, GSI Helmholtzzentrum f\"ur Schwerionenforschung GmbH, Darmstadt, Germany\\
$^{97}$ Saga University, Saga, Japan\\
$^{98}$ Saha Institute of Nuclear Physics, Homi Bhabha National Institute, Kolkata, India\\
$^{99}$ School of Physics and Astronomy, University of Birmingham, Birmingham, United Kingdom\\
$^{100}$ Secci\'{o}n F\'{\i}sica, Departamento de Ciencias, Pontificia Universidad Cat\'{o}lica del Per\'{u}, Lima, Peru\\
$^{101}$ Stefan Meyer Institut f\"{u}r Subatomare Physik (SMI), Vienna, Austria\\
$^{102}$ SUBATECH, IMT Atlantique, Nantes Universit\'{e}, CNRS-IN2P3, Nantes, France\\
$^{103}$ Sungkyunkwan University, Suwon City, Republic of Korea\\
$^{104}$ Suranaree University of Technology, Nakhon Ratchasima, Thailand\\
$^{105}$ Technical University of Ko\v{s}ice, Ko\v{s}ice, Slovak Republic\\
$^{106}$ The Henryk Niewodniczanski Institute of Nuclear Physics, Polish Academy of Sciences, Cracow, Poland\\
$^{107}$ The University of Texas at Austin, Austin, Texas, United States\\
$^{108}$ Universidad Aut\'{o}noma de Sinaloa, Culiac\'{a}n, Mexico\\
$^{109}$ Universidade de S\~{a}o Paulo (USP), S\~{a}o Paulo, Brazil\\
$^{110}$ Universidade Estadual de Campinas (UNICAMP), Campinas, Brazil\\
$^{111}$ Universidade Federal do ABC, Santo Andre, Brazil\\
$^{112}$ Universitatea Nationala de Stiinta si Tehnologie Politehnica Bucuresti, Bucharest, Romania\\
$^{113}$ University of Derby, Derby, United Kingdom\\
$^{114}$ University of Houston, Houston, Texas, United States\\
$^{115}$ University of Jyv\"{a}skyl\"{a}, Jyv\"{a}skyl\"{a}, Finland\\
$^{116}$ University of Kansas, Lawrence, Kansas, United States\\
$^{117}$ University of Liverpool, Liverpool, United Kingdom\\
$^{118}$ University of Science and Technology of China, Hefei, China\\
$^{119}$ University of South-Eastern Norway, Kongsberg, Norway\\
$^{120}$ University of Tennessee, Knoxville, Tennessee, United States\\
$^{121}$ University of the Witwatersrand, Johannesburg, South Africa\\
$^{122}$ University of Tokyo, Tokyo, Japan\\
$^{123}$ University of Tsukuba, Tsukuba, Japan\\
$^{124}$ Universit\"{a}t M\"{u}nster, Institut f\"{u}r Kernphysik, M\"{u}nster, Germany\\
$^{125}$ Universit\'{e} Clermont Auvergne, CNRS/IN2P3, LPC, Clermont-Ferrand, France\\
$^{126}$ Universit\'{e} de Lyon, CNRS/IN2P3, Institut de Physique des 2 Infinis de Lyon, Lyon, France\\
$^{127}$ Universit\'{e} de Strasbourg, CNRS, IPHC UMR 7178, F-67000 Strasbourg, France, Strasbourg, France\\
$^{128}$ Universit\'{e} Paris-Saclay, Centre d'Etudes de Saclay (CEA), IRFU, D\'{e}partment de Physique Nucl\'{e}aire (DPhN), Saclay, France\\
$^{129}$ Universit\'{e}  Paris-Saclay, CNRS/IN2P3, IJCLab, Orsay, France\\
$^{130}$ Universit\`{a} degli Studi di Foggia, Foggia, Italy\\
$^{131}$ Universit\`{a} del Piemonte Orientale, Vercelli, Italy\\
$^{132}$ Universit\`{a} di Brescia, Brescia, Italy\\
$^{133}$ Variable Energy Cyclotron Centre, Homi Bhabha National Institute, Kolkata, India\\
$^{134}$ Warsaw University of Technology, Warsaw, Poland\\
$^{135}$ Wayne State University, Detroit, Michigan, United States\\
$^{136}$ Yale University, New Haven, Connecticut, United States\\
$^{137}$ Yildiz Technical University, Istanbul, Turkey\\
$^{138}$ Yonsei University, Seoul, Republic of Korea\\
$^{139}$ Affiliated with an institute formerly covered by a cooperation agreement with CERN\\
$^{140}$ Affiliated with an international laboratory covered by a cooperation agreement with CERN.\\

\end{flushleft} 
  
\end{document}